\documentclass{article}
\usepackage{amsmath, amsthm}
\usepackage{amssymb}
\newtheorem{proposition}{Proposition}
\newtheorem{theorem}{Theorem}
\newtheorem{remark}{Remark}
\newtheorem{definition}{Definition}

\usepackage[margin=1in,dvips]{geometry}
\usepackage{bbm}
\usepackage{graphics}

\def\ub {\underline{u}}
\def\th {\theta}
\def\Lb {\underline{L}}
\def\Hb {\underline{H}}
\def\chib {\underline{\chi}}
\def\chih {\hat{\chi}}
\def\chibh {\hat{\underline{\chi}}}
\def\omegab {\underline{\omega}}
\def\etab {\underline{\eta}}
\def\betab {\underline{\beta}}
\def\alphab {\underline{\alpha}}

\def\hot{\widehat{\otimes}}

\def\thb {\underline{\theta}}
\def\t {\tilde}
\def\a {\alpha}
\def\b {\beta}
\def\ab {\alphab}
\def\bb {\betab}
\def\nab {\nabla}

\renewcommand{\div}{\mbox{div }}
\newcommand{\curl}{\mbox{curl }}
\newcommand{\trchb}{\mbox{tr} \chib}
\def\trch{\mbox{tr}\chi}

\newcommand{\eps}{{\epsilon} \mkern-8mu /\,}
\newcommand{\Ls}{{\mathcal L} \mkern-10mu /\,}

\title{On the Local Existence for the Characteristic Initial Value Problem in General Relativity}
\author{Jonathan Luk}

\begin{document}

\maketitle

\begin{abstract}
Given a truncated incoming null cone and a truncated outgoing null cone intersecting at a two sphere $S$ with smooth characteristic initial data, a theorem of Rendall shows that the vacuum Einstein equations can be solved in a small neighborhood of $S$ in the future of $S$. We show that in fact the vacuum Einstein equations can be solved in a neighborhood in the future of the cones, as long as the constraint equations are initially satisfied on the null cones. The proof is based on energy type estimates and relies heavily on the null structure of the Einstein equations in the double null foliation.
\end{abstract}

\section{Introduction}

In this paper, we study the characteristic initial value problem of the vacuum Einstein equation. We work in the setting of an outgoing null cone $H_0$ intersecting with an incoming null cone $\Hb_0$ at a two sphere $S_{0,0}$ (See Figure 1).

\begin{figure}[htbp]
\begin{center}
 
\input{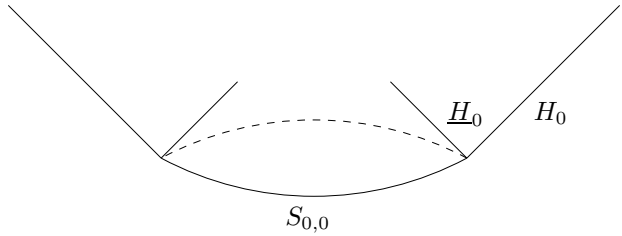}
 
\caption{Basic Setup}
\end{center}
\end{figure}

It was shown by Rendall \cite{Rendall} that for smooth characteristic initial data prescribed on $H_0$ and $\Hb_0$, the Einstein equations can be solved in a neighborhood of the intersecting sphere to the future of the two null cones as shown by the shaded region in the following figure.

\begin{figure}[htbp]
\begin{center}
 
\input{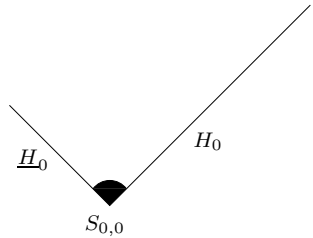}
 
\caption{Region of Existence in Rendall's Theorem}
\end{center}
\end{figure}

A very natural question which we learned in a talk of Rendall at MSRI in 2009 is whether one has local existence in a neighborhood of the two null cones, instead of only in a neighborhood of the intersecting sphere. One obstruction is that it is not always possible to solve the constraint equations on the initial characteristic hypersurfaces away from the intersecting sphere. Nevertheless, this question can be studied for characteristic initial data set satisfying the constraint equations. We prove that if the constraint equations are satisfied on $H_0$ for $0\leq \ub\leq I_1$ and on $\Hb_0$ for $0\leq u\leq I_2$ for some $I_1,I_2>0$, then there exists $\epsilon$ such that the Einstein equations can be solved for $\{0\leq \ub\leq I_1\}\cap\{0\leq u\leq \epsilon\}$ and $\{0\leq u\leq I_2\}\cap\{0\leq \ub\leq \epsilon\}$ as indicated in the following figure. Moreover, $\epsilon$ depends only on the size (measured in appropriate norms) of the characteristic initial data. 

\begin{figure}[htbp]
\begin{center}
 
\input{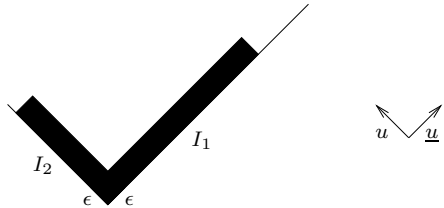}
 
\caption{Improved Region of Existence}
\end{center}
\end{figure}

\begin{theorem}\label{main}
Given regular characteristic initial data that satisfy the constraint equations, there exists a regular solution to the Einstein equations (unique in the double null foliation) in a neighborhood to the future of the null cones. Moreover, the size of the neighborhood can be made to depend only on the size of the initial data.
\end{theorem}

A precise formulation of characteristic initial value problem and the theorem, as well as the definition of the notion of regular initial data can be found in the next section. Notice that while we work with smooth characteristic initial data, the size of the neighborhood in the above theorem depends only on the $H^5$ norm of the metric. Moreover, using a standard approximation procedure, we can construct an $H^4$ solution to the Einstein equations given $H^5$ characteristic initial data. An argument to improve the regularity of this theorem is sketched in Section \ref{regularity}.

Rendall's original work covers also the case where the initial characteristic data are prescribed on two intersecting null hyperplanes. Our method can also be extended to this case to show that if the constraints are satisfied, the Einstein equations can be solved in a neighborhood to the future of the null hyperplanes. Recently, Choquet-Bruhat, Chrusciel and Martin-Garcia \cite{CBCMG} studied the case where the initial characteristic data are prescribed on a cone. They constructed data satisfying the constraints and solved the Einstein equations near the vertex. 
 
One reason for studying the characteristic initial value problem is that contrary to the usual Cauchy problem in general relativity, the constraint equations for the characteristic initial value problem are of the type of transport equations. They are therefore much easier to analyze compared to the constraint equations for the Cauchy problem. This can be used to construct special initial data set for which the dynamics can be understood. An important example is the recent monumental work of Christodoulou on the formation of trapped surfaces \cite{Chr}. 

While Rendall's theorem holds for general quasilinear wave equations, our theorem only holds when a special ``null structure'' is present in the equation. In Section \ref{semilinear}, we will show that the ideas in this paper can be modified to treat the case of a semilinear wave equation with a null condition in $3+1$ Minkowski space. In Section \ref{counterexample}, we will also indicate how a corresponding statement fails for a general semilinear wave equation not satisfying the null condition. This thus shows the importance of the structure of the Einstein's equations in the double null foliation.

We now indicate the main ideas in the proof. We foliate the spacetime with outgoing null hypersurfaces $H_u$ and incoming null hypersurfaces $\Hb_{\ub}$ such that the characteristic initial value is prescribed on $H_0$ and $\Hb_0$. Then define an appropriately normalized null frame $\{e_1,e_2,e_3,e_4\}$ such that $e_3$ is tangent to $\Hb_{\ub}$, $e_4$ is tangent to $H_u$, and $\{e_1, e_2\}$ is a frame tangent to the 2-spheres where $H_u$ and $\Hb_{\ub}$ intersect. Define the Ricci coefficients $\psi=g(D_{e_\mu}e_{\nu}, e_{\sigma})$ and the null curvature components $\Psi=R(e_\mu,e\nu,e_\sigma,e_\delta)$ with respect to this frame. We would like to show that for characteristic initial value of size $\sim 1$ satisfying the constraint equations for $0\leq \ub\leq I$ (for some $I>0$) and $u=0$, there is a sufficiently small $\epsilon$ such that we can prove estimates for the curvature and Ricci coefficients of the spacetime in the region $0\leq \ub\leq I$ and $0\leq u\leq \epsilon$. This would then allow us to prove that existence of solution in this region.

Following the general strategy in \cite{CK}, \cite{KN}, \cite{Chr} etc., we prove the desired estimates in two steps. 
In the first step, we assume the ($L^2$) bounds on the (derivatives of the) curvature component and try to prove estimates for the Ricci coefficients. 
$$\nabla_3\psi=\Psi+\psi\psi,$$
$$\nabla_4\psi=\Psi+\psi\psi.$$
The difficulty is to integrate the nonlinear terms. For the $\nabla_3$ equations, we can take advantage of the small $\epsilon$ length and show that the Ricci coefficients have norms very close to their initial value. However, there are two components, namely, $\eta$ and $\omegab$, that do not satisfy a $\nabla_3$ equation but satisfy only a $\nabla_4$ equation. The key observation is that for the $\nabla_4\eta$ equation, all the nonlinear terms are of the form that at least one of the factors can be estimated by a $\nabla_3$ equation. In other words, the terms $\eta^2$, $\omegab^2$ and $\eta\omegab$ do not appear. Since using the $\nabla_3$ equations we have already established that the norms for at least one of the factors have norms very close to their initial value, the equation becomes essentially linear. Once $\eta$ is estimated, we can move to the $\nabla_4\omegab$ equation, for which $\omegab^2$ does not appear in the nonlinear term and we can therefore integrate and prove estimates for all the Ricci coefficients.

The second step is the energy estimates for the curvature components. Neglecting derivatives, these are $L^2$ estimates of the form 
$$\int_{H_u} \Psi^2+\int_{\Hb_{\ub}} \Psi^2 \leq \mathcal R_0+\int_{D_{u,\ub}}\psi\Psi^2.$$
If either one of the curvature components $\Psi$ in the error term $\int_{D_{u,\ub}}\psi\Psi^2$ can be controlled on $H_u$, we can integrate along the $u$ direction to gain a power of $\epsilon$. However, the component $\alphab$ cannot be controlled on $H_u$ but can only be controlled on $\Hb_{\ub}$. In order to control the term with $\alphab^2$, we notice that the Ricci coefficient coupling to it satisfies a $\nabla_3$ equation and thus can be controlled by a constant depending only on the initial data. Hence this term can be controlled using the Gronwall inequality.

In the next section, we will lay out the basic setup, define the double null foliation and write down the equations in our setting. In section \ref{local}, we will prove Rendall's result \cite{Rendall} in the double null foliation using his reduction of the characteristic initial value problem to the Cauchy problem. In section \ref{geometry}, we prove the basic estimates that are necessary for the estimates for the Ricci coefficients and the curvature components. In section \ref{estimates}, we prove the estimates for the Ricci coefficients and the curvature components. Then in section \ref{lastslice}, we show that this allows us to prove our main theorem. We conclude in section \ref{discussion} with some comparisons with semilinear wave equations in $3+1$ Minkowski space and some discussions on how regularity can be improved for our main theorem.

{\bf Acknowledgements.} The author thanks his advisor Igor Rodnianski for many enlightening discussions, as well as many helpful comments for improving the manuscript.

\section{Basic Setup}

\subsection{Canonical Coordinate System}

Our basic setup is depicted in the following diagram.

\begin{figure}[htbp]
\begin{center}
 
\input{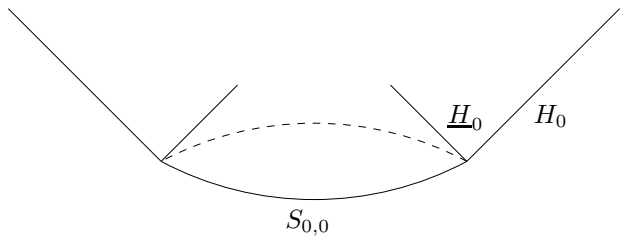}
 
\caption{Basic Setup}
\end{center}
\end{figure}

The two hypersurfaces $H_0$ and $\Hb_0$ are prescribed to be null. The intersection of the two hypersurfaces is a spacelike 2-sphere which we denote as $S_{0,0}$. We consider a spacetime in the future of the two null cones. In this spacetime, we consider optical functions $u$ and $\ub$ satisfying the eikonal equation
$$g^{\mu\nu}\partial_\mu u\partial_\nu u=0,\quad g^{\mu\nu}\partial_\mu\ub\partial_\nu \ub=0,$$
where $u=0$ on $H_0$ and $\ub=0$ on $\Hb_0$.

Let
$$L'^\mu=-2g^{\mu\nu}\partial_\nu u,\quad \Lb'^\mu=-2g^{\mu\nu}\partial_\nu \ub.$$ 
Define
$$2\Omega^{-2}=-g(L',\Lb').$$
Define
$$e_3=\Omega\Lb'\mbox{, }e_4=\Omega L'.$$
and
$$\Lb=\Omega^2\Lb'\mbox{, }L=\Omega^2 L'.$$

We will denote the level sets of $u$ as $H_u$ and the level sets of $\ub$ and $\Hb_{\ub}$. By virtue of the Eikonal equations, $H_u$ and $\Hb_{\ub}$ are null hypersurface. We will use the notation $H_u(\ub',\ub'')$ to denote the part of the hypersurface $H_u$ with $\ub'\leq \ub\leq \ub''$. We will also use $\Hb_{\ub}(u',u'')$ in the obvious way. Notice that the sets defined by fixed values of $(u,\ub)$ are 2-spheres. We denote such spheres by $S_{u,\ub}$. They are intersections of the hypersurfaces $H_u$ and $\Hb_{\ub}$. We will also denote the region $\displaystyle\bigcup_{0\leq u'\leq u, 0\leq \ub'\leq\ub}S_{u,\ub'}$ by $D_{u,\ub}$.

We introduce a coordinate system $(u,\ub,\th^1,\th^2)$ as follows:
On the sphere $S_{0,0}$, define a coordinate system $(\th^1,\th^2)$ in a coordinate patch $U$ for the sphere. On $H_0$, we define the coordinate system $(\ub,\th^1,\th^2)$ such that $\frac{\partial}{\partial \ub}=L$ is tangent to the null geodesics that generate $H_0$. We then define the coordinate system in the full spacetime by letting $u$ and $\ub$ to be solutions to the Eikonal equations as above and define $\th^1, \th^2$ by
$$\Lb (\th^A)=0.$$ 
For each coordinate patch $U$, a system of coordinates $(u,\ub,\th^1,\th^2)$ is thus defined on $D_U$, where $D_U$ is defined to be the image of first applying the diffeomorphism generated by $\Lb$ on $\Hb_0$, then applying the diffeomorphism generated by $L$.

In these coordinates, we have 
$$e_3=\Omega^{-1}\frac{\partial}{\partial u},\quad e_4=\Omega^{-1}\left(\frac{\partial}{\partial \ub}+b^A\frac{\partial}{\partial \th^A}\right),$$
for some $b^A$ such that $b^A=0$ on $H_0$.
In these coordinates, the metric is
$$g=-2\Omega^2(du\otimes d\ub+d\ub\otimes du)+\gamma_{AB}(d\th^A-b^Adu)\otimes (d\th^B-b^Bdu).$$ 
Here, $\gamma_{AB}$ is the restriction of the spacetime metric to the tangent space of $S_{u,\ub}$. We can choose $u$ and $\ub$ in such a way that $\Omega=1$ on the initial hypersurfaces $H_0$ and $\Hb_0$. From this point onwards, we will make this assumption. This can be thought of as a normalization condition for $u$ and $\ub$. When we make the assertion that the spacetime exists up to $u\leq \epsilon$, we always take into account this normalization of $u$.

We will also use a system of coordinates $(u,\ub,\thb^1,\thb^2)$ which is defined in a similar way as the coordinate system $(u,\ub,\th^1,\th^2)$, except for reversing the roles of $L$ and $\Lb$ in the definition. We will use $(u,\ub,\th^1,\th^2)$ as a coordinate system near $H_0$ and $(u,\ub,\thb^1,\thb^2)$ as a coordinate system near $\Hb_0$.
\subsection{The Equations}
Given a   2-sphere  $S_{u,\ub}$  and  $(e_A)_{A=1,2}$  an arbitrary  frame tangent to it  we define
 \begin{equation}
\begin{split}
&\chi_{AB}=g(D_A e_4,e_B),\, \,\, \quad \chib_{AB}=g(D_A e_3,e_B),\\
&\eta_A=-\frac 12 g(D_3 e_A,e_4),\quad \etab_A=-\frac 12 g(D_4 e_A,e_3)\\
&\omega=-\frac 14 g(D_4 e_3,e_4),\quad\,\,\, \omegab=-\frac 14 g(D_3 e_4,e_3),\\
&\zeta_A=\frac 1 2 g(D_A e_4,e_3)
\end{split}
\end{equation}
where $D_A=D_{e_{(A)}}$. We also introduce the  null curvature components,
 \begin{equation}
\begin{split}
\a_{AB}&=R(e_A, e_4, e_B, e_4),\quad \, \,\,   \ab_{AB}=R(e_A, e_3, e_B, e_3),\\
\b_A&= \frac 1 2 R(e_A,  e_4, e_3, e_4) ,\quad \bb_A =\frac 1 2 R(e_A,  e_3,  e_3, e_4),\\
\rho&=\frac 1 4 R(e_4,e_3, e_4,  e_3),\quad \sigma=\frac 1 4  \,^*R(e_4,e_3, e_4,  e_3)
\end{split}
\end{equation}
Here $\, ^*R$ denotes the Hodge dual of $R$.  We denote by $\nab$ the 
induced covariant derivative operator on $S_{u,\ub}$ and by $\nab_3$, $\nab_4$
the projections to $S_{u,\ub}$ of the covariant derivatives $D_3$, $D_4$, see
precise definitions in \cite{KN}. Moreover, we define $\phi^{(1)}\cdot\phi^{(2)}$ to be an arbitrary contraction of the tensor product of $\phi^{(1)}$ and $\phi^{(2)}$ with respect to the metric $\gamma$ and also
$$(\phi^{(1)}\hot\phi^{(2)})_{AB}:=\phi^{(1)}_A\phi^{(2)}_B+\phi^{(1)}_B\phi^{(2)}_A-\delta_{AB}(\phi^{(1)}\cdot\phi^{(2)}) \quad\mbox{for one forms $\phi^{(1)}_A$, $\phi^{(2)}_A$,}$$
$$(\phi^{(1)}\wedge\phi^{(2)})_{AB}:=\eps^{AB}(\gamma^{-1})^{CD}\phi^{(1)}_{AC}\phi^{(2)}_{BD}\quad\mbox{for symmetric two tensors $\phi^{(1)}_{AB}$, $\phi^{(2)}_{AB}$}.$$
Define the divergence and curl of totally symmetric tensors to be
$$(\div\phi)_{A_1...A_r}:=\nabla^B\phi_{BA_1...A_r},$$
$$(\curl\phi)_{A_1...A_r}:=\eps^{BC}\nabla_B\phi_{CA_1...A_r},$$
where $\eps$ is the volume form associated to the metric $\gamma$.
Define also the trace to be
$$(\mbox{tr}\phi)_{A_1...A_{r-1}}:=(\gamma^{-1})^{BC}\phi_{BCA_1...A_{r-1}}.$$
Also, denote by $^*$ the Hodge dual on $S_{u,\ub}$.
Observe that,
\begin{equation}
\begin{split}
&\omega=-\frac 12 \nab_4 (\log\Omega),\qquad \omegab=-\frac 12 \nab_3 (\log\Omega),\\
&\eta_A=\zeta_A +\nab_A (\log\Omega),\quad \etab_A=-\zeta_A+\nab_A (\log\Omega)
\end{split}
\end{equation}
We separate the trace and traceless part of $\chi$ and $\chib$. Let $\chih$ and $\chibh$ be the traceless parts of $\chi$ and $\chib$ respectively. Then $\chi$ and $\chib$ satisfy the following null structure equations:
\begin{equation}
\label{null.str1}
\begin{split}
\nab_4 \trch+\frac 12 (\trch)^2&=-|\chih|^2-2\omega \trch\\
\nab_4\chih+\trch \chih&=-2 \omega \chih-\alpha\\
\nab_3 \trchb+\frac 12 (\trchb)^2&=-2\omegab \trchb-|\chibh|^2\\
\nab_3\chibh + \trchb\,  \chibh&= -2\omegab \chibh -\alphab\\
\nab_4 \trchb+\frac1 2 \trch \trchb &=2\omega \trchb +2\rho- \chih\cdot\chibh +2\div \etab +2|\etab|^2\\
\nab_4\chibh +\frac 1 2 \trch \chibh&=\nab\widehat{\otimes} \etab+2\omega \chibh-\frac 12 \trchb \chih +\etab\widehat{\otimes} \etab\\
\nab_3 \trch+\frac1 2 \trchb \trch &=2\omegab \trch+2\rho- \chih\cdot\chibh+2\div \eta+2|\eta|^2\\
\nab_3\chih+\frac 1 2 \trchb \chih&=\nab\widehat{\otimes} \eta+2\omegab \chih-\frac 12 \trch \chibh +\eta\widehat{\otimes} \eta
\end{split}
\end{equation}
The other Ricci coefficients satisfy the following null structure equations:
\begin{equation}
\label{null.str2}
\begin{split}
\nabla_4\eta&=-\chi\cdot(\eta-\etab)-\b\\
\nabla_3\etab &=-\chib\cdot (\etab-\eta)+\bb\\
\nabla_4\omegab&=2\omega\omegab+\frac 34 |\eta-\etab|^2-\frac 14 (\eta-\etab)\cdot (\eta+\etab)-
\frac 18 |\eta+\etab|^2+\frac 12 \rho\\
\nabla_3\omega&=2\omega\omegab+\frac 34 |\eta-\etab|^2+\frac 14 (\eta-\etab)\cdot (\eta+\etab)- \frac 18 |\eta+\etab|^2+\frac 12 \rho\\
\end{split}
\end{equation}
The Ricci coefficients also satisfy the following constraint equations
\begin{equation}
\label{null.str3}
\begin{split}
\div\chih&=\frac 12 \nabla \trch - \frac 12 (\eta-\etab)\cdot (\chih -\frac 1 2 \trch) -\beta,\\
\div\chibh&=\frac 12 \nabla \trchb + \frac 12 (\eta-\etab)\cdot (\chibh-\frac 1 2   \trchb) +\betab\\
\curl\eta &=-\curl\etab=\sigma +\frac 1 2\chibh \wedge\chih\\
K&=-\rho+\frac 1 2 \chih\cdot\chibh-\frac 1 4 \trch \trchb
\end{split}
\end{equation}
with $K$ the Gauss curvature of the surfaces $S$.
The null curvature components satisfy the following null Bianchi equations:
\begin{equation}
\label{eq:null.Bianchi}
\begin{split}
&\nab_3\alpha+\frac 12 \trchb \alpha=\nabla\hot \beta+ 4\omegab\alpha-3(\chih\rho+^*\chih\sigma)+
(\zeta+4\eta)\hot\beta,\\
&\nab_4\beta+2\trch\beta = \div\alpha - 2\omega\beta +  \eta \alpha,\\
&\nab_3\beta+\trchb\beta=\nabla\rho + 2\omegab \beta +^*\nabla\sigma +2\chih\cdot\betab+3(\eta\rho+^*\eta\sigma),\\
&\nab_4\sigma+\frac 32\trch\sigma=-\div^*\beta+\frac 12\chibh\cdot ^*\alpha-\zeta\cdot^*\beta-2\etab\cdot
^*\beta,\\
&\nab_3\sigma+\frac 32\trchb\sigma=-\div ^*\betab+\frac 12\chih\cdot ^*\alphab-\zeta\cdot ^*\betab-2\eta\cdot 
^*\betab,\\
&\nab_4\rho+\frac 32\trch\rho=\div\beta-\frac 12\chibh\cdot\alpha+\zeta\cdot\beta+2\etab\cdot\beta,\\
&\nab_3\rho+\frac 32\trchb\rho=-\div\betab- \frac 12\chih\cdot\alphab+\zeta\cdot\betab-2\eta\cdot\betab,\\
&\nab_4\betab+\trch\betab=-\nabla\rho +^*\nabla\sigma+ 2\omega\betab +2\chibh\cdot\beta-3(\etab\rho-^*\etab\sigma),\\
&\nab_3\betab+2\trchb\betab=-\div\alphab-2\omegab\betab+\etab \cdot\alphab,\\
&\nab_4\alphab+\frac 12 \trch\alphab=-\nabla\hot \betab+ 4\omega\alphab-3(\chibh\rho-^*\chibh\sigma)+
(\zeta-4\etab)\hot \betab
\end{split}
\end{equation}

In the sequel, we will use capital Latin letters $A\in \{1,2\}$ for indices on the spheres $S_{u,\ub}$ and Greek letters $\mu\in\{1,2,3,4\}$ for indices in the whole spacetime.

It will be useful in the following to use a schematic notation. We will use $\phi$ to denote an arbitrary tensorfield. We will denote Ricci coefficients by $\psi$ and null curvature components by $\Psi$. Unless otherwise stated, $\psi$ will denote an arbitrary Ricci coefficients and $\Psi$ can denote an arbitrary null curvature components. We will simply write $\psi\psi$ (or $\psi\Psi$, etc.) to denote contractions using the metric $\gamma$. When we use this notation, the exact way that the tensors are contracted is irrelevant to the argument. Moreover, when using this schematic notation, we will neglect all constant factors.

\subsection{Initial Data}

On the initial characteristic hypersurface, $\gamma$, $\chi$ and $\chib$ have to satisfy the equations
\begin{equation}\label{con1}
\Ls_L \gamma=2\chi,
\end{equation}
\begin{equation}\label{con2}
\Ls_{\Lb}\gamma=2\chib,
\end{equation}
\begin{equation}\label{con3}
\Ls_L \trch= -\frac 12 (\trch)^2-|\chih|_\gamma^2
\end{equation}
\begin{equation}\label{con4}
\Ls_{\Lb} \trchb= -\frac 12 (\trchb)^2-|\chibh|_\gamma^2
\end{equation}
Here $\Ls$ denotes the restriction of the Lie derivative to $TS_{u,\ub}$. This is a notion intrinsic to the null hypersurfaces.

\begin{definition}
By an initial data set, we refer to a quadruple $(\gamma,\chi,\chib,\zeta)$ such that $\gamma$ is a positive definite symmetric covariant two tensorfield on $S_{0,0}$, $\chi$ is a symmetric covariant two tensorfield on $S_{0,\ub}$ for $\ub\in [0,I_1]$, $\chib$ is a symmetric covariant two tensorfield on $S_{u,0}$ for $u\in [0,I_2]$, and $\zeta$ is a covariant one tensorfield on $S_{0,0}$.
\end{definition}

\begin{definition}
We say that an initial data set $(\gamma,\chi,\chib,\zeta)$ is regular if (\ref{con1}), (\ref{con3}) are satisfied on $H_0$ and (\ref{con2}), (\ref{con4}) are satisfied on $\Hb_0$, $\gamma$ is positive definite and that the quantities $\gamma$, $\chi$, $\chib$, $\zeta$ are $C^\infty$.
\end{definition}

On the initial outgoing hypersurface $H_0$ we prescribe the conformal class of the metric $\hat{\gamma}_{AB}$ satisfying in coordinates $\sqrt{\det \hat{\gamma}_{AB}}=1$. Similarly, on the initial incoming hypersurface $\Hb_0$ we prescribe we prescribe the conformal class of the metric $\hat{\gamma}_{AB}$ satisfying in coordinates $\sqrt{\det \hat{\gamma}_{AB}}=1$. Prescribe also $\gamma_{AB}$, $\zeta_A$, $\trch$ and $\trchb$ on the two sphere $S_{0,0}$. On $S_{0,0}$, since $\hat{\gamma}_{AB}$ and $\gamma_{AB}$ are in the same conformal class, 
$$\gamma_{AB}=\Phi^2 \hat{\gamma}_{AB}.$$
By (\ref{con1}), we know that in the canonical coordinates,
$$\frac{\partial}{\partial \ub}\gamma_{AB}=2\chi_{AB}=2\chih_{AB}+\trch \gamma_{AB}.$$
On the other hand
$$\frac{\partial}{\partial \ub}\gamma_{AB}=\Phi^2\frac{\partial}{\partial \ub}\hat\gamma_{AB}+2\Phi\frac{\partial\Phi}{\partial\ub}\hat\gamma_{AB}.$$
Since we know that 
$$({\hat\gamma}^{-1})^{AB}\frac{\partial}{\partial \ub}\hat\gamma_{AB}=\frac{\partial}{\partial \ub}\log\det\hat\gamma=0,$$
we can identify
$$\chih_{AB}=\frac{1}{2}\Phi^2\frac{\partial}{\partial \ub}\hat\gamma_{AB},$$
and
$$\trch=\frac{2}{\Phi}\frac{\partial\Phi}{\partial\ub}.$$
Notice also that 
$$|\chih|_\gamma^2=\frac 14 ({\hat\gamma}^{-1})^{AC}({\hat\gamma}^{-1})^{BD}\frac{\partial}{\partial \ub}\hat\gamma_{AB}\frac{\partial}{\partial \ub}\hat\gamma_{CD}$$
depends only on $\hat\gamma$. Thus, (\ref{con3}) can be re-written as
$$\frac{\partial^2\Phi}{\partial \ub^2}+\frac 18 ({\hat\gamma}^{-1})^{AC}({\hat\gamma}^{-1})^{BD}\frac{\partial}{\partial \ub}\hat\gamma_{AB}\frac{\partial}{\partial \ub}\hat\gamma_{CD}=0.$$
This ordinary differential equation can be solved with the appropriate initial data. Locally, we also know that $\Phi\neq 0$, and thus the solution is regular. In general, we cannot guarantee that the solution to this ODE is regular up to $\ub=1$, but we will only consider data that satisfy this assumption. The equations (\ref{con2}), (\ref{con4}) on $\Hb_0$ can be solved in a similar fashion.

\subsection{Integration and Norms}

We define the integration on $S_{u,\ub}$ and $D_{u,\ub}$ in the natural way: Let $U$ be a coordinate patch on $S_{0,0}$ and $p_U$ be a partition of unity in $D_U$ such that $p_U$ is supported in $D_U$. Given a function $\phi$, we define the integration by the volume form of the induced metric on $S_{u,\ub}$:
$$\int_{S_{u,\ub}} \phi :=\sum_U \int_{-\infty}^{\infty}\int_{-\infty}^{\infty}\phi p_U\sqrt{\det\gamma}d\th^1 d\th^2.$$
On $D_{u,\ub}$, we define integration using the volume form of the spacetime metric
\begin{equation*}
\begin{split}
\int_{D_{u,\ub}} \phi :=&\sum_U \int_0^u\int_0^{\ub}\int_{-\infty}^{\infty}\int_{-\infty}^{\infty}\phi p_U\sqrt{-\det g}d\th^1 d\th^2d\ub' du\\
=&2\sum_U \int_0^u\int_0^{\ub}\int_{-\infty}^{\infty}\int_{-\infty}^{\infty}\phi p_U\Omega^2\sqrt{-\det \gamma}d\th^1 d\th^2d\ub' du.
\end{split}
\end{equation*}

There are no canonical volume forms on $H_u$ and $\Hb_{\ub}$. We will define integration by
$$\int_{H_{u}(0,\ub)} \phi :=\sum_U \int_0^{\ub}\int_{-\infty}^{\infty}\int_{-\infty}^{\infty}\phi2 p_U\Omega\sqrt{\det\gamma}d\th^1 d\th^2d\ub,$$
and
$$\int_{H_{\ub}(0,u)} \phi :=\sum_U \int_0^u\int_{-\infty}^{\infty}\int_{-\infty}^{\infty}\phi2p_U\Omega\sqrt{\det\gamma}d\th^1 d\th^2du.$$
We also write 
$$\int_{H_u}=\int_{H_u(0,I)},$$
and
$$\int_{\Hb_{\ub}}=\int_{\Hb_{\ub}(0,\epsilon)}.$$
We define these norms so that the quantities are integrated ``near'' $H_0$. We will prove our main theorem near $H_0$. All the estimates near $\Hb_0$ follow with identical arguments and one can define the corresponding norms in the obvious manner.

With these definitions of integration, we can define the norms that we will use. Let $\phi$ be a tensorfield. For $1\leq p<\infty$, define
$$||\phi||_{L^p(S_{u,\ub})}^p:=\int_{S_{u,\ub}} <\phi,\phi>_\gamma^{p/2},$$
$$||\phi||_{L^p(H_u)}^p:=\int_{H_{u}} <\phi,\phi>_\gamma^{p/2},$$
$$||\phi||_{L^p(\Hb_{\ub})}^p:=\int_{\Hb_{\ub}} <\phi,\phi>_\gamma^{p/2}.$$
Define also the $L^\infty$ norm by
$$||\phi||_{L^\infty(S_{u,\ub})}:=\sup_{\th\in S_{u,\ub}} <\phi,\phi>_\gamma^{1/2}(\th).$$
It is easy to note that all quantities will be the same if we define the norms and integrations instead using the $(u,\ub,\thb^1,\thb^2)$ coordinates.
\subsection{Statement of the Theorem}
We will prove the following theorem:
\begin{theorem}\label{mainthm}
Given regular initial data on $H_0$ for $0\leq \ub\leq I$. Then there exists $\epsilon$ such that a smooth solution to the vacuum Einstein equations exist in the region such that $0\leq \ub\leq I$ and $0\leq u\leq \epsilon$. This solution is unique in the canonical coordinates.
Moreover, $\epsilon$ can be chosen to depend only on
\begin{equation*}
\begin{split}
\mathcal G_0:=&\sup_{S\subset H_0, \mbox{ }S\subset \Hb_0}\sup_U(\sup_{A,B=1,2} \gamma_{AB}+ \det\gamma+(\inf \det\gamma)^{-1})+I,\\
\mathcal O_0:=&\sup_{S\subset H_0,\mbox{ }S\subset \Hb_0}\sup_{\psi\in\{\chih,\trch,\omega,\eta,\etab,\chibh,\trchb,\omegab\}}\max\{1,\sum_{i=0}^{3}||\nabla^i\psi||_{L^2(S)},\sum_{i=0}^{2}||\nabla^i\psi||_{L^4(S)},\sum_{i=0}^1||\nabla^i\psi||_{L^\infty(S)}\},\\
\mathcal R_0:=&
\sum_{i=0}^{3}\left(\sup_{\Psi\in\{\alpha,\beta,\rho,\sigma,\betab\}}||\nabla^i\Psi||_{L^2(H_0)}+\sup_{\Psi\in\{\beta,\rho,\sigma,\betab,\alphab\}}||\nabla^i\Psi||_{L^2(\Hb_0)}\right)\\
&+\sup_{S\subset H_0,\mbox{ }S\subset \Hb_0}\sup_{\Psi\in\{\alpha,\beta,\rho,\sigma,\betab,\alphab\}}\max\{1,\sum_{i=0}^{2}||\nabla^i\Psi||_{L^2(S)},\sum_{i=0}^1||\nabla^i\Psi||_{L^4(S)}\}.
\end{split}
\end{equation*}
Moreover, in this spacetime, 
\begin{equation*}
\begin{split}
&\sup_{u,\ub}\sup_{\psi\in\{\chih,\trch,\omega,\eta,\etab,\chibh,\trchb,\omegab\}}\max\{\sum_{i=0}^{3}||\nabla^i\psi||_{L^2(S_{u,\ub})},\sum_{i=0}^{2}||\nabla^i\psi||_{L^4(S_{u,\ub})},\sum_{i=0}^1||\nabla^i\psi||_{L^\infty(S_{u,\ub})}\}\\
&+\sum_{i=0}^3(\sup_u\sup_{\Psi\in\{\alpha,\beta,\rho,\sigma,\betab\}}||\nabla^i\Psi||_{L^2(H_u)}+\sup_{\ub}\sup_{\Psi\in\{\beta,\rho,\sigma,\betab,\alphab\}}||\nabla^i\Psi||_{L^2(\Hb_{\ub})})\\
\leq &C(\mathcal O_0,\mathcal R_0,\mathcal G_0).
\end{split}
\end{equation*}

\end{theorem}
With the same argument, we can also prove a similar statement near $\Hb_0$. This would then imply our main theorem as stated in the introduction.
\begin{remark}
Notice that the value of $\mathcal G_0$ does depend on the choice of coordinates.
\end{remark}

\section{Rendall's Theorem}\label{local}

In this section, we repeat the proof of Rendall's Theorem \cite{Rendall}. The main goal of this section is to show that the local existence theorem of Rendall holds for the characteristic initial data prescribed in the double null foliation. Moreover, we show the uniqueness of solutions in the canonical coordinates.

The proof in \cite{Rendall} goes in two steps. Firstly, a local existence theorem is proved for general quasilinear wave equations. Then, this theorem is applied to the Einstein equations. 
Choose the wave coordinates so that 
$$\Gamma^\mu=g^{\nu\sigma}\Gamma^\mu_{\nu\sigma}=0.$$
It is well-known that using the wave coordinates, the Einstein equations are equivalent to the so-called reduced Einstein equations
$$\tilde{R}_{\mu\nu}=R_{\mu\nu}+g_{\sigma(\mu}\Gamma^{\sigma}_{,\nu)}=0,$$ 
which can be written as a system of quasilinear wave equations of $g_{\mu\nu}$. It is therefore sufficient to show that the condition for the wave coordinates is satisfied for the solution. In \cite{Rendall}, it is shown that given the conformal class of the metrics on the spheres on the initial characteristic hypersurface, one can choose a coordinate system and prescribe the other components of the metric in this coordinate system such that $\Gamma^\mu=0$ on the initial characteristic hypersurfaces. Since $\Gamma^\mu$ also satisfies a system of quasilinear wave equations, by uniqueness, $\Gamma^\mu$ is identically zero. The main goal of this section is thus to show that for our prescribed characteristic initial value, we can also introduce a coordinate system so that $\Gamma^\mu=0$ on the initial characteristic hypersurfaces, as in \cite{Rendall}.

For completeness, we cite a particular case of Rendall's Theorem:
\begin{theorem}[Rendall \cite{Rendall}]\label{qlwethm}
Consider a quasilinear wave equation 
\begin{equation}\label{qlwe}
g^{\mu\nu}(\phi)\partial_{\mu\nu}^2\phi=F(\phi,\partial\phi),
\end{equation}
where $g$ and $F$ are smooth in its variables, with smooth initial data $\phi$ on two null hypersurfaces $H_0$ and $\Hb_0$. (Initial data are prescribed in a way that $H_0$ and $\Hb_0$ are null.) Suppose that $H_0$ and $\Hb_0$ intersect in a topological 2-sphere $S_{0,0}$. Suppose moreover that all derivatives of $\phi$ are continuous up to $S_{0,0}$. Then, there exists a small neighborhood of $S_{0,0}$ in the future of $S_{0,0}$ such that a unique solution to (\ref{qlwe}) exists.
\end{theorem}
Using this theorem, we can prove local existence for the Einstein equations in a small neighborhood of $S_{0,0}$ in the future of $S_{0,0}$ in the canonical coordinates.
\begin{theorem}
Given a regular initial data set, there exists a small neighborhood to the future of $S_{0,0}$ such that the Einstein equations can be solved.
\end{theorem}
\begin{proof}
We first focus on the outgoing hypersurface $H_0$. The incoming hypersurface $\Hb_0$ can be treated analogously. On $H_0$, consider a coordinate patch for the canonical coordinate system. Rename the coordinates 
$$x^1=\th^1, x^2=\th^2, \ub=x^4.$$
Then let
$g_{AB}=\gamma_{AB}$ and $g_{44}=g_{4A}=0$. By (\ref{con1}) and (\ref{con3}), we have
$$g_{AB,4}=2\chi_{AB},$$
and 
$$\frac{1}{2}(g^{AB}g_{AB,4})_{,4}+\frac{1}{4}g^{AB}{  }_{,4}g_{AB,4}=0.$$
It is helpful to note that since $g_{4A}=0$ and $g_{44}=0$, we must have
$$g^{3A}=0,$$
$$g^{33}=0,$$
$$g^{34}g_{34}=1,\mbox{ and}$$
$$g^{33}{  }_{,3}=-g^{34}g^{34}g_{44,3}.$$
We would like to prescribe $g_{33}, g_{34}, g_{3A}$ on $H_0$ such that the wave coordinate constraints are satisfied and we can solve the reduced Einstein equations.

We solve for $f_3$ from the ODE
\begin{equation}\label{f3}
\frac{\partial f_3}{\partial x^4}=\frac{1}{2}g^{AB}g_{AB,4}f_3, 
\end{equation}
with the initial condition 
$$f_3=1 \mbox{ on }S_{0,0}.$$
Prescribe
$$g_{34}=-2f_3. $$
Notice that since $f_3\sim 1$, we have $g_{34}\sim -2$ near $S_{0,0}$. Thus, near $S_{0,0}$, we have $g_{34}\neq 0$ and $g^{34}\neq 0$.
In wave coordinates, we would have the condition
$$\Gamma^3=0,$$
which, in coordinates, reads
\begin{equation}\label{Gamma3}
g_{44,3}=\frac{1}{2}g^{AB}g_{AB,4}g_{34}.
\end{equation}
We will for now assume that this is the value of $g_{44,3}$. Once we have proved the existence of the spacetime, we will show that this condition is indeed satisfied in the spacetime that we have constructed. Notice that $g_{34}$ and $g_{44,3}$ that we prescribe,
$$R_{44}=\frac{1}{2}(g^{AB}g_{AB,4})_{,4}+\frac{1}{4}g^{AB}{  }_{,4}g_{AB,4}-\frac{1}{4}g^{34}g^{AB}g_{AB,4}(2g_{34,4}-g_{44,3})=0.$$
We note also that since we have prescribed $g_{34}$, we can compute $g_{34,A}$.

On $S_{0,0}$, we prescribe $g_{3A}$ by the following procedure. First, we write the prescribed $\zeta_A$ in coordinates. Then we require on $S_{0,0}$ that
$$g_{3A,4}-g_{4A,3}=4\zeta_A-g_{34,A}.$$
Moreover, we can write down the wave coordinate condition $\Gamma^A=0$, which in coordinates reads
\begin{equation}\label{GammaA}
\Gamma^A=\frac{1}{2}g^{34}g^{AB}(g_{B3,4}+g_{B4,3})+\frac{1}{2}g^{34}g^{A4}g_{44,3}+...=0,
\end{equation}
where $...$ represents terms that can be computed from the metric components that we know so far. Now these two conditions give four linearly independent linear equations on $g_{31,4}, g_{32,4}, g_{41,3}, g_{42,3}$ and hence we can solve for all of these on $S_{0,0}$. Now we proceed to prescribe $g_{3A,4}$ on $H_0$. We consider the equation  $R_{4A}=0$ in coordinates, using $g_{44,3}=\frac{1}{2}g^{AB}g_{AB,4}g_{34}$:
\begin{equation}\label{R4A}
\begin{split}
0=R_{4A}=&\frac{1}{2}(g^{34}(g_{3A,4}-g_{4A,3}))_{,4}+\frac{1}{2}(g^{4B}g_{AB,4})_{,4}+\frac{1}{2}g^{3B}{  }_{,3}g_{AB,4}\\
&-\frac{1}{2}g^{CD}g_{CD,4}(g^{34}(g_{3A.4}-g_{4A,3})+g^{4B}g_{AB,4})-\frac{1}{2}g^{BC}g^{34}g_{AC,4}g_{4C,B}+...,
\end{split}
\end{equation}
where $...$ denotes terms involving the metric components that we have already prescribed. We note that $g^{4B}$ is linear in $g_{3C}$ with coefficients that are smooth functions of the metric components that we have already prescribed. We also note that $g^{3B}{  }_{,3}$ is a linear function in $g_{4A,3}$ in the same sense. If we impose $\Gamma^A=0$ and $\Gamma^A{ }_{,4}=0$, then
$$R_{4A}=(g^{34}g_{3A,4})_{,4}+K_{4A}(g_{3B}, g_{3B,4},1),$$
where $K_{4A}$ is a linear function in $g_{3B}, g_{3B,4},1$ with coefficients being smooth functions of the metric components that we have already prescribed. (\ref{R4A}) is thus a second order ordinary differential equation for $g_{3A}$. We solve this equation with the initial condition on $S_{0,0}$ $g_{3A}=0$ and $g_{3A,4}$ as found above. Once $g_{3A}$ is retrieved, we also have $g_{4A,3}$ from the condition (\ref{GammaA}). As for $g_{44,3}$, $g_{4A,3}$ is not a function that we can prescribe to solve the reduced Einstein equations. We will nevertheless check that $g_{4A,3}$ is indeed this function in the spacetime that we construct by checking that (\ref{GammaA}) holds.

We now prescribe $g_{33}$ on $H_0$. We would impose the condition
$$\Gamma^4=0,$$
which in coordinates reads
\begin{equation}\label{Gamma4}
g^{34}g_{33,4}-g^{AB}g_{AB,3}+g^{44}g_{44,3}+...=0,
\end{equation}
where as before, $...$ denotes terms that we have already prescribed, i.e., terms involving $g_{AB},g_{A4},g_{34},g_{3A},g_{44,3},g_{4A,3}$ and their derivatives along $H_0$. Notice that $g^{44}$ is a linear function in $g_{33}$ depending on the terms that we have already prescribed..
Now consider the conditions $R_{34}=0$ and $R_{AB}=0$. In coordinates, we have
\begin{equation*}
\begin{split}
0=R_{AB}=&-\frac{1}{2}(g^{34}g_{AB,4})_{,3}-\frac{1}{2}(g^{34}g_{AB,3})_{,4}-\frac{1}{2}(g^{44}g_{AB,4})_{,4}\\
&-\frac{1}{2}g^{34}g^{CD}(g_{AC,3}g_{BD,4}+g_{AC,4}g_{BD,3})-\frac{1}{2}g^{44}g^{CD}g_{AC,4}g_{BD,4}\\
&+\frac{1}{4}g^{34}g_{AB,4}(g^{44}g_{44,3}+2g^{34}g_{34,3})+\frac{1}{2}g^{34}g_{AB,3}(g^{34}g_{34,4}+g^{CD}g_{CD,4})+...
\end{split}
\end{equation*}
Hence, we can write
\begin{equation*}
\begin{split}
0=R_{AB}=&-g^{34}g_{AB,34}+\t K_{AB}(g_{33},g_{33,4},g_{CD,3},g_{34,4},1),
\end{split}
\end{equation*}
where $\t K_{AB}$ is linear in $g_{33},g_{33,4},g_{CD,3},g_{34,4},1$ with coefficients being quantities that we have already prescribed. We now substitute in (\ref{Gamma4}) to get
\begin{equation}\label{RAB}
\begin{split}
R_{AB}=&-g^{34}g_{AB,34}+K_{AB}(g_{33},g_{CD,3},g_{34,4},1)=0,
\end{split}
\end{equation}
For $R_{34}=0$, we have, in coordinates,
\begin{equation*}
\begin{split}
0=R_{34}=&-\frac{1}{2}(g^{34}g_{33,4})_{,4}+\frac{1}{2}(g^{34}(2g_{34,4}-g_{44,3}))_{,3}+\frac{1}{2}(g^{AB}g_{AB,4})_{,3}-\frac{1}{2}g^{44}{  }_{,4}g_{44,3}\\
&+\frac{1}{4}g^{AC}g^{BD}g_{BC,4}g_{AD,3}-\frac{1}{4}g^{AB}g_{AB,4}(g^{34}g_{33,4}+g^{44}g_{44,3})\\
&-\frac{1}{4}g^{34}g_{44,3}(g^{34}g_{33,4}+g^{44}g_{44,3})-\frac{1}{4}g^{34}g^{AB}g_{44,3}g_{AB,3}+...
\end{split}
\end{equation*}
Note that $g^{44}$ is linear in $g_{33}$ and $g^{34}{  }_{,3}$ is linear in $g_{34,3}$ and $g_{AB,3}$. Suppose we have the condition 
$$\Gamma^4=0,\quad \Gamma^3{ }_{,3}=0.$$
Then
\begin{equation*}
\begin{split}
R_{34}=&-\frac{1}{2}(g^{34}g_{33,4})_{,4}+(g^{34}g_{34,4})_{,3}+\t K_{34}(g_{33},g_{33,4},g_{AB,3},g_{34,4},1)\\
=&\frac{1}{2}(g^{AB}g_{AB,3})_{,4}+(g^{34}g_{34,4})_{,3}+\t K_{34}(g_{33},g_{33,4},g_{AB,3},g_{34,4},1).
\end{split}
\end{equation*}
where $\t K_{34}$ is a function, linear in $g_{33}, g_{33,4}, g_{AB,3},g_{34,4},1$ with coefficients depending on the previously prescribed quantities. Now, substituting in (\ref{Gamma4}) and (\ref{RAB}), we have
\begin{equation}\label{R34}
\begin{split}
R_{34}=(g^{34}g_{34,4})_{,3}+ K_{34}(g_{33},g_{AB,3},g_{34,4},1)=0.
\end{split}
\end{equation}
We now have a coupled system of five linear ordinary differential equations (\ref{Gamma4}), (\ref{RAB}) and (\ref{R34}) which are first order in $g_{33}$, $g_{AB,3}$ and $g_{34,3}$. The initial conditions for the ordinary differential equations are dictated by continuity on $\Hb_0$. Thus we want $g_{33}=0$, $g_{34,3}=0$ identically on $S_{0,0}$ and $g_{AB,3}$ to be given by the prescribed initial value.

Notice that assuming all $g_{\mu\nu}$ and $g_{AB,3}, g_{44,3}, g_{4A,3}, g_{34,3}$ as above, we also have $R_{44,3}=0$. To see this, we use the contracted Bianchi identity
$$\nabla^\mu G_{\mu 4}=0,$$
where $G_{\mu\nu}=R_{\mu\nu}-\frac{1}{2}g_{\mu\nu}R$ with $R$ being the scalar curvature. First, notice that all the terms in this identity are terms that we have prescribed in the above procedure (i.e., $g_{3A,3}$, $g_{33,3}$ and terms that involve 2 derivatives in the $3$ direction do not appear). Then notice that since $g^{3A}=g^{33}=0$, $R_{44}=R_{4A}=R_{34}=R_{AB}=0$ implies $R=0$. Now, using $R_{4\mu}=0$ and $R=0$ on $H_0$, we know that the only potentially non-vanishing term is the $3$ derivative of $G_{\mu 4}$. Thus, we have
$$0=g^{34}\nabla_3 G_{44}=g^{34}(R_{44,3}-\frac{1}{2}(g_{44}R)_{,3}+2\Gamma^\mu_{34}G_{4\mu})=g^{34}R_{44,3}.$$
The non-vanishing of $g^{34}$ in the small neighborhood of $S_{0,0}$ thus gives 
$$R_{44,3}=0.$$

Now we have given $g_{\mu\nu}$ on $H_0$ and $\Hb_0$. By Rendall's Theorem \ref{qlwethm} on the local existence for quasilinear wave equations, we can solve the reduced Einstein equations. 
$$\t R_{\mu,\nu} =R_{\mu\nu}+g_{\sigma(\mu}\Gamma^\sigma{ }_{,\nu)}=0,$$
which is a system of quasilinear wave equations.

We need to show that this solution to the reduced Einstein equations is indeed a solution to the Einstein equations. By the construction of the data, we have prescribed all derivatives of $g$ on $S_{0,0}$ such that $\Gamma^\mu=0$ and $R_{\mu\nu}=0$. We would like to show that in fact $\Gamma^\mu=0$ on $H_0$ and $\Hb_0$. As before, we will consider the situation on $H_0$. $\Hb_0$ can be treated analogously. Consider the equation
$$\tilde{R}_{44}=R_{44}+g_{34}\Gamma^{3}{ }_{,4}=0.$$
On $H_0$, $g_{\mu\nu}, g_{\mu\nu,A}, g_{\mu\nu,4}, g_{\mu\nu,4A}, g_{\mu\nu,44}$ are as prescribed. Hence $\tilde{R}_{44}=0$ is a first order ordinary equation for $g_{44,3}$. Since $\Gamma^3=0$ and $R_{44}=0$ satisfy the initial conditions and give a solution to the ODE, by uniqueness of solutions, it must be the case that $\Gamma^3=0$ and $R_{44}=0$ on $H_0$. Thus we have moreover shown that $g_{44,3}=\frac{1}{2}g^{AB}g_{AB,4}g_{34},$ as indicated before. Now consider
$$\tilde{R}_{4A}=R_{4A}+\frac{1}{2}g_{34}\Gamma^{3}{ }_{,A}+\frac{1}{2}g_{3A}\Gamma^{3}{ }_{,4}+\frac{1}{2}g_{AB}\Gamma^{B}{ }_{,4}=0.$$
The only terms that are not determined by the initial data are $g_{4A,3}, g_{4A,34}, g_{44,3}, g_{44,3A}, g_{44,34}$. We know, nevertheless that $g_{44,3}, g_{44,3A}, g_{44,34}$ can be determined by the condition $\Gamma^3=0$. Hence, $\tilde{R}_{4A}=0$ is a system of first order ODEs for $g_{4A,3}$. Notice that the highest order term $g_{4A,34}$ appears in both $R_{4A}$ and $\frac{1}{2}g_{3A}\Gamma^3{ }_{,4}$. We need to make sure that the coefficient in the highest order term in not degenerate. Indeed the term is
$$-\frac{1}{2}g^{34}g_{4A,34}+\frac{1}{4}g^{34}g_{4A,34}=-\frac{1}{4}g^{34}g_{4A,34}.$$
We know that $g^{34}\neq 0$ near $S_{0,0}$ and thus we can solve for $g_{4A,3}$ using $\tilde{R}_{4A}=0$. Since $g_{4A,3}$ given by $\Gamma^A=0$ is a solution, by uniqueness, we must have $R_{4A}=0$ and $\Gamma^A=0$ on $H_0$. It remains to show that $\Gamma^4=0$. To do so, we need to consider the equations $\tilde{R}_{AB}=\tilde{R}_{34}=\tilde{R}_{44,3}=0$. Using $\Gamma^3=\Gamma^A=0$,
$$\tilde{R}_{AB}=R_{AB}=0,$$
The only terms that are not prescribed as initial data and have not been determined by $\tilde{R}_{44}=\tilde{R}_{A4}=0$ are
$$\tilde{R}_{AB}=-\frac{1}{2}(g^{34}g_{AB,4})_{,3}-\frac{1}{2}(g^{34}g_{AB,3})_{,4}-\frac{1}{2}g^{34}g^{CD}(g_{AC,3}g_{BD,4}+g_{AC,4}g_{BD,3})+\frac{1}{2}g^{34}g^{34}g_{AB,4}g_{34,3}+...=0.$$
We can write
\begin{equation}\label{tRAB}
\tilde{R}_{AB}=-g^{34}g_{AB,34}+H_{AB}(g_{CD,3},g_{34,3},1)=0,
\end{equation}
where $H_{AB}(g_{CD,3},g_{34,3},1)$ is linear in $g_{CD,3},g_{34,3},1$.
Next, using $\Gamma^3=\Gamma^A=0$, we have
$$\tilde{R}_{34}=R_{34}+\frac{1}{2}g_{34}\Gamma^4{ }_{,4}=0.$$
Again, we note that terms that are not prescribed as initial data and have not been determined by $\tilde{R}_{44}=\tilde{R}_{A4}=0$:
\begin{equation*}
\begin{split}
\tilde{R}_{34}=&(g^{34}g_{34,4})_{,3}-\frac{1}{2}(g^{34}g_{44,3})_{,3}+\frac{1}{2}(g^{AB}g_{AB,4})_{,3}\\
&+\frac{1}{4}g^{AC}g^{BD}g_{BC,4}g_{AD,3}-\frac{1}{4}g^{34}g^{AB}g_{44,3}g_{AB,3}+\frac{1}{4}(-g^{AB}g_{AB,34}+g^{44}g_{44,34})+...=0.
\end{split}
\end{equation*}
Thus, we can write
\begin{equation*}
\tilde{R}_{34}=g^{34}g_{34,34}+\tilde{H}_{34}(g_{AB,34}, g_{44,33},g_{CD,3},g_{34,3},1)=0.
\end{equation*}
We can substitute the equation (\ref{tRAB}) for $\tilde{R}_{AB}=0$ to replace the term $g_{AB,34}$ to get
\begin{equation}\label{tR34}
\tilde{R}_{34}=g^{34}g_{34,34}+{H}_{34}(g_{44,33},g_{CD,3},g_{34,3},1)=0.
\end{equation}
Finally, using $\Gamma^3=\Gamma^A=0$, we also have
\begin{equation*}
\tilde{R}_{44,3}=R_{44,3}+g_{34}\Gamma^{3}{ }_{,34}=0.
\end{equation*}
Again, we note that terms that are not prescribed as initial data and have not been determined by $\tilde{R}_{44}=\tilde{R}_{34}=0$:
$$\tilde{R}_{44,3}=R_{44,3}+g_{34}\Gamma^{3}{ }_{,34}=0.$$
$$\tilde{R}_{44,3}=-\frac{1}{4}g^{34}g^{AB}(g_{AB,34}(2g_{34,4}-g_{44,3})+g_{AB,4}(2g_{34,34}-g_{44,33}))+\frac{1}{4}g^{34}g_{44,334}...$$
Thus, we can write
$$\tilde{R}_{44,3}=\frac{1}{4}g^{34}g_{44,334}+\tilde{H}_{443}(g_{34,34},g_{AB,34},g_{44,33},g_{CD,3},g_{34,3},1)=0,$$
where $\tilde{H}$ is linear in $g_{34,34},g_{AB,34},g_{44,33},g_{CD,3},g_{34,3},1$ with coefficients that have been prescribed as initial data or have been determined by $\tilde{R}_{44}=\tilde{R}_{34}=0$. Substituting (\ref{tRAB}) and (\ref{tR34}), we get
\begin{equation}\label{tR443}
\tilde{R}_{44,3}=\frac{1}{4}g^{34}g_{44,334}+{H}_{443}(g_{44,33},g_{CD,3},g_{34,3},1)=0.
\end{equation}
Therefore, by (\ref{tRAB}), (\ref{tR34}) and (\ref{tR443}), we have a coupled system on first order ODEs for $g_{AB,3}, g_{34,3}, g_{44,33}$. Uniqueness would now demand that $\Gamma^4=\Gamma^3{ }_{,3}=0, R_{34}=R_{AB}=R_{44,3}=0$ on $H_0$.

We have thus established that $\Gamma^\mu=0$ on $H_0$ and, by symmetry, $\Hb_0$. It is well-known that $\Gamma^\mu$ satisfies a system of quasilinear wave equation. Therefore, by the uniqueness part in Theorem \ref{qlwethm}, $\Gamma^\mu=0$ in the spacetime, whenever it exists. Hence, the solution to the reduced Einstein equations is indeed a solution to the Einstein equations. Thus to conclude the existence result, it remains to show that the solution satisfies the prescribed initial data. To do so, we change coordinates back to our original gauge. Solve for 
$$(g^{-1})^{\mu\nu}\partial_\mu u \partial_\nu u=0, \quad (g^{-1})^{\mu\nu}\partial_\mu \ub \partial_\nu \ub=0$$
with the conditions $u=x^3$, $\ub=x^4$ on $H_0$ and $\Hb_0$. Now define $\th^1$, $\th^2$ by
$$\frac{\partial \th^1}{\partial u}=\frac{\partial \th^2}{\partial u}=0.$$
We notice that the Jacobian on each point on $S_{0,0}$ is the identity. Thus in a neighborhood of $S_{0,0}$, $(\th^1, \th^2, u, \ub)$ forms a coordinate system. By virtue of $u, \ub$ being solutions to the Eikonal equation, it is clear that
$$g_{uu}=g_{\ub \ub}=0.$$
Clearly, $\gamma$ and $\zeta$ are as prescribed on $S_{0,0}$. It remains to show that $g_{u \ub}=-2$ on $H_0$ and $\Hb_0$ and that $\chi$ and $\chib$ are as prescribed on $H_0$ and $\Hb_0$ respectively. We will first compute $g_{u \ub}$ on $H_0$. The case for $\Hb_0$ can be treated analogously. Consider
$$0=\frac{\partial}{\partial x^3}((g^{-1})^{\mu\nu}\partial_\mu u \partial_\nu u)=g^{34}\frac{\partial u}{\partial x^3}(-g^{34}g_{44,3}\frac{\partial u}{\partial x^3}+\frac{\partial^2 u}{\partial x^3\partial x^4}).$$
Using (\ref{Gamma3}), we thus have
$$\frac{\partial }{\partial x^4}\frac{\partial u}{\partial x^3}=\frac{1}{2}g^{AB}g_{AB,4}\frac{\partial u}{\partial x^3}.$$
Moreover, we know by continuity and the fact that $u=x^3$ on $\Hb_0$ that $\frac{\partial u}{\partial x^3}=1$ on $S_{0,0}$. In other words, $\frac{\partial u}{\partial x^3}$ satisfies (\ref{f3}) and has the same initial condition as $f_3$. Thus, $\frac{\partial u}{\partial x^3}=f_3$ on $H_0$. Now,
$$g_{u \ub}=(\frac{\partial u}{\partial x^3})^{-1}g_{34}=-2(f_3)^{-1}f_3=-2.$$
Finally, we compute that 
$$\chi_{AB}=\frac{1}{2}g_{AB,4}\quad\mbox{on $H_0$},\quad\mbox{and }\chib_{AB}=\frac{1}{2}g_{AB,3}\quad\mbox{on $\Hb_0$},$$
as desired.
\end{proof}
\begin{theorem}
The solution is unique in the canonical coordinates. 
\end{theorem}
\begin{proof}
Give a solution to the Einstein equations with given initial data, solve the linear wave equation
$$\Box_g x^\mu=0,$$
with $(x^1, x^2,x^3,x^4)$ being the original coordinate functions on $H_0$ and $\Hb_0$. This gives a change of coordinates in a neighborhood in the future to $S_{0,0}$ since the Jacobian is the identity matrix on $H_0$ and $\Hb_0$. Since the condition for wave coordinates is satisfied, in this new coordinate system we must have
\begin{equation}\label{wavegauge}
\Gamma^\mu=0.
\end{equation}
Moreover, since the spacetime is a solution to the Einstein equations,
\begin{equation}\label{EE}
R_{\mu\nu}=0.
\end{equation}
By the proof of the previous theorem, (\ref{wavegauge}) and (\ref{EE}) together uniquely determine all components of $g_{\mu\nu}$ in this new coordinate system. Now since (\ref{wavegauge}) is satisfied, the Einstein equations are equivalent to the reduced Einstein equations. Hence, the metric components satisfy the reduced Einstein equations, which is a quasilinear wave equation. Hence, the metric is uniquely determined.
\end{proof}

\section{Basic Estimates}\label{geometry}
In this section, we would like to assume the appropriate boundedness of the Ricci coefficients and would like to obtain three types of basic estimates. First, we would like to control the metric components in the canonical coordinates. Second, we would like to show the equivalence of norms using the control of the metric components. Third, we would like to prove basic estimates for Sobolev embedding, $L^2$ elliptic estimates and estimates for the covariant transport equations.

Some estimates that we derive depend on the coordinates we choose. We will derive the estimates near $H_0$ using the coordinate system $(u,\ub,\th^1,\th^2)$. The estimates near $\Hb_0$ will be similar if we use the coordinate system $(u,\ub,\th^1,\th^2)$.

All estimates will be proved under the following bootstrap assumption:
\begin{equation}\label{BA}
||(\chih,\chibh,\trch,\trchb,\zeta,\omega,\omegab)||_{L^\infty(S_{u,\ub})}\leq \Delta_0,
\end{equation}
where $\Delta_0$ is a large constant to be chosen later. Notice that while the choice of $\epsilon$ depends on $\Delta_0$, all the estimates are independent of $\Delta_0$.

\subsection{Estimates for Metric Components}\label{metric}
We first show that we can control $\Omega$ in $\mathcal D$ with this bootstrap assumption:
\begin{proposition}\label{Omega}
For $\epsilon$ small enough depending on initial data and $\Delta_0$, there exists $C$ depending only on initial data such that
$$\Omega,\Omega^{-1}\leq C.$$
\end{proposition}
\begin{proof}
\[
 \omega=-\frac{1}{2}\nabla_3\log\Omega=\frac{1}{2}\Omega\nabla_3\Omega^{-1}=\frac{1}{2}\frac{\partial}{\partial u}\Omega^{-1}.
\]
Now both $\omega$ and $\Omega$ are scalars and therefore the $L^\infty$ norm is independent of the metric. We can show that $\Omega^{-1}$ is close to the corresponding value of $\Omega^{-1}$ on a sphere that is on the $H_0$. More precisely, fix $\ub$. Then 
$$||\Omega^{-1}-1||_{L^\infty(S_{u,\ub})}\leq C\int_0^{u}||\omega||_{L^\infty(S_{u',\ub})}du\leq C\Delta_0\epsilon.$$
This implies both the estimates for $\Omega$ and $\Omega^{-1}$ for sufficiently small $\epsilon$.
\end{proof}

We then show that we can control $\gamma$ in $\mathcal D$ with the bootstrap assumption:
\begin{proposition}\label{gamma}
Consider a coordinate patch $U$ on $S_{0,0}$ and define $U_{0,\ub}$ to be a coordinate patch on $S_{0,\ub}$ given by the one-parameter diffeomorphism generated by $L$. Define $U_{u,\ub}$ to be the image of $U_{0,\ub}$ under the one-parameter diffeomorphism generated by $\Lb$. Define also $D_U=\bigcup_{0\leq u\leq I,0\leq \ub\leq \epsilon} U_{u,\ub}$. We require $\det\gamma$ to be bounded above and below on $U_{0,\ub}$. By the assumption of the regular initial data, each point on $S_{0,\ub}$ lies in such a $U_{0,\ub}$.  For $\epsilon$ small enough depending on initial data and $\Delta_0$, there exists $C$ and $c$ depending only on initial data such that the following pointwise bounds for $\gamma$ in $\mathcal D_U$ hold:
$$c\leq \det\gamma\leq C. $$
Moreover, in $D_U$,
$$|\gamma_{AB}|,|(\gamma^{-1})^{AB}|\leq C.$$
\end{proposition}
\begin{proof}
The first variation formula states that
$$\Ls_L\gamma=2\Omega\chi.$$
In coordinates, this means
$$\frac{\partial}{\partial \ub}\gamma_{AB}=2\Omega\chi_{AB}.$$
From this we derive that 
$$\frac{\partial}{\partial \ub}\log(\det\gamma)=\Omega\trch.$$
Define $\gamma_0(u,\ub,\th^1,\th^2)=\gamma(0,\ub,\th^1,\th^2)$. 
$$|\det\gamma-\det(\gamma_0)|\leq C\int_0^{\ub}|\trch|d\ub'\leq C\Delta_0\epsilon.$$
We thus know that the $\det \gamma$ is bounded above and below. Let $\Lambda$ be the larger eigenvalue of $\gamma$. Clearly,
\begin{equation}\label{La}
\Lambda\leq C\sup_{A,B=1,2}\gamma,
\end{equation}
and 
$$\sum_{A,B=1,2}|\chi_{AB}|^2\leq C\Lambda ||\chi||_{L^\infty(S_{u,\ub})}.$$
Then
$$|\gamma_{AB}-(\gamma_0)_{AB}|\leq C\int_0^{\ub}|\chi_{AB}|d\ub'\leq C\Lambda\Delta_0\epsilon.$$
Using the upper bound (\ref{La}), we thus obtain the upper bound for $|\gamma_{AB}|$. The upper bound for $|(\gamma^{-1})^{AB}|$ follows from the upper bound for $|\gamma_{AB}|$ and the lower bound for $\det\gamma$.
\end{proof}

A consequence of the previous Proposition is an estimate on the surface area of each two sphere $S_{u,\ub}$.
\begin{proposition}\label{area}
$$\sup_{u,\ub}|\mbox{Area}(S_{u,\ub})-\mbox{Area}(S_{0,\ub})|\leq C\Delta_0\epsilon.$$
\end{proposition}
\begin{proof}
This follows from the fact that $\sqrt{\det\gamma}$ is pointwise only slightly perturbed if $\epsilon$ is chosen to be appropriately small.
\end{proof}
With the estimate on the volume form, we can now show that the $L^p$ norms defined with respect to the metric and the $L^p$ norms defined with respect to the coordinate system are equivalent.
\begin{proposition}\label{eqnorm}
Given a covariant tensor $\phi_{A_1...A_r}$ on $S_{u,\ub}$, we have
$$\int_{S_{u,\ub}} <\phi,\phi>_{\gamma}^{p/2} \sim \sum_{i=1}^r\sum_{A_i=1,2}\iint |\phi_{A_1...A_r}|^p \sqrt{\det\gamma} d\th^1 d\th^2.$$
\end{proposition}
We can also control $b$ under the bootstrap assumption, thus controlling the full spacetime metric: 
\begin{proposition}\label{b}
In the canonical coordinates,
$$|b^A|\leq C\Delta_0\epsilon.$$
\end{proposition}
\begin{proof}
$b^A$ satisfies the equation
$$\frac{\partial b^A}{\partial u}=-4\Omega^2\zeta^A.$$
This can be derived from 
$$[L,\Lb]=-\frac{\partial b^A}{\partial u}\frac{\partial}{\partial \th^A}.$$
Now, integrating and using Proposition \ref{eqnorm} gives the result.
\end{proof}

\subsection{Estimates for Transport Equations}
We need to use the null structure equations and the null Bianchi equations to obtain estimates for the Ricci coefficients and the null curvature components respectively. In order to use the equations, we need a way to obtain estimates from the null transport type equations. This can be achieved under the assumption of bounded of $\trch$ and $\trchb$.
\begin{proposition}\label{transport}
Assume
$$\sup_{u,\ub}||\trch,\trchb||_{L^\infty(S_{u,\ub})}\leq 4\mathcal O_0.$$
Then there exists $\epsilon_0=\epsilon_0(\mathcal O_0)$ such that for all $\epsilon \leq \epsilon_0$ and for every $1\leq p<\infty$, we have
\[
 ||\phi||_{L^p(S_{u,\ub})}\leq C(\mathcal O_0, I)\left(||\phi||_{L^p(S_{u,\ub'})}+\int_{\ub'}^{\ub} ||\nabla_4\phi||_{L^p(S_{u,\ub''})}d{\ub''}\right)
\]
\[
 ||\phi||_{L^p(S_{u,\ub})}\leq 2(||\phi||_{L^p(S_{u',\ub})}+\int_{u'}^{u} ||\nabla_3\phi||_{L^p(S_{u'',\ub})}d{u''}).
\]
\end{proposition}

\begin{proof}

The following identity holds for any scalar $f$:
\[
 \frac{d}{d\ub}\int_{\mathcal S_{u,\ub}} f=\int_{\mathcal S_{u,\ub}} \left(\frac{df}{d\ub}+\Omega \trch f\right)=\int_{\mathcal S_{u,\ub}} \Omega\left(e_4(f)+ \trch f\right).
\]
Similarly, we have
\[
 \frac{d}{du}\int_{\mathcal S_{u,\ub}} f=\int_{\mathcal S_{u,\ub}} \Omega\left(e_3(f)+ \trchb f\right).
\]
This can be proved by using a different coordinate system $(u,\ub,\thb^1,\thb^2)$
Hence, taking $f=|\phi|_{\gamma}^2$, we have
\[
 ||\phi||^2_{L^2(S_{u,\ub})}=||\phi||^2_{L^2(S_{u,\ub'})}+\int_{\ub'}^{\ub}\int_{S_{u,\ub''}} 2\Omega\left(<\phi,\nabla_4\phi>_\gamma+ \frac{1}{2}\trch |\phi|^2_{\gamma}\right)d{\ub''}
\]
\[
 ||\phi||^2_{L^2(S_{u,\ub})}=||\phi||^2_{L^2(S_{u',\ub})}+\int_{u'}^{u}\int_{S_{u'',\ub}} 2\Omega\left(<\phi,\nabla_3\phi>_\gamma+ \frac{1}{2}\trchb |\phi|^2_{\gamma}\right)d{u''}
\]
The Proposition is proved using Cauchy-Schwarz on the sphere and the $L^\infty$ bounds for $\Omega$ and $\trch$ ($\trchb$) which are provided by Proposition \ref{Omega} and the assumption respectively. For the $L^4$ estimates, take $f=|\phi|_{\gamma}^4$, and we have
\begin{equation}\label{L4transport}
 ||\phi||^4_{L^4(S_{u,\ub})}=||\phi||^4_{L^2(S_{u,\ub'})}+\int_{\ub'}^{\ub}\int_{S_{u,\ub''}} 2\Omega|\phi|_{\gamma}^2\left(<\phi,\nabla_4\phi>_\gamma+ \frac{1}{2}\trch |\phi|^2_{\gamma}\right)d{\ub''}
\end{equation}
\[
 ||\phi||^4_{L^4(S_{u,\ub})}=||\phi||^2_{L^2(S_{u',\ub})}+\int_{u'}^{u}\int_{S_{u'',\ub}} 2\Omega|\phi|_{\gamma}^2\left(<\phi,\nabla_3\phi>_\gamma+ \frac{1}{2}\trchb |\phi|^2_{\gamma}\right)d{u''}
\]
Again, we conclude using Cauchy-Schwarz on the sphere and the $L^\infty$ bounds for $\Omega$ and $\trch$ ($\trchb$).
\end{proof}
The above estimates also hold for $p=\infty$:
\begin{proposition}\label{transportinfty}
There exists $\epsilon_0=\epsilon_0(\mathcal O_0)$ such that for all $\epsilon \leq \epsilon_0$, we have
\[
 ||\phi||_{L^\infty(S_{u,\ub})}\leq C(\mathcal O_0, I)\left(||\phi||_{L^\infty(S_{u,\ub'})}+\int_{\ub'}^{\ub} ||\nabla_4\phi||_{L^\infty(S_{u,\ub''})}d{\ub''}\right)
\]
\[
 ||\phi||_{L^\infty(S_{u,\ub})}\leq 2(||\phi||_{L^\infty(S_{u',\ub})}+\int_{u'}^{u} ||\nabla_3\phi||_{L^\infty(S_{u'',\ub})}d{u''}).
\]
\end{proposition}
\begin{proof}
This follows simply from integrating along the integral curves of $L$ and $\Lb$, and the estimate on $\Omega$ in Proposition \ref{Omega}.
\end{proof}
Using this type of estimates we can control the (derivatives of the) Ricci coefficients assuming the appropriate bounds on the (derivatives of the) null curvature components. 

\subsection{Sobolev Embedding}
From the estimate of the metric, we can also show the Sobolev Embedding Theorems on the two spheres $S_{u,\ub}$. 
\begin{proposition}\label{L4}
There exists $\epsilon_0=\epsilon_0(\Delta_0,\mathcal G_0)$ such that as long as $\epsilon\leq \epsilon_0$, we have
$$||\phi||_{L^4(S_{u,\ub})}\leq C(\mathcal G_0)\sum_{i=0}^1||\nabla^i\phi||_{L^2(S_{u,\ub})}. $$
\end{proposition}
\begin{proof}
We first prove this for scalars. Since we already have a coordinate system, we only need to estimate the volume form. By Proposition \ref{gamma}, however, we know that the volume form is bounded above and below. Thus, the proposition holds for scalars. Now, for $\phi$ being a tensor, let $f=\sqrt{|\phi|_{\gamma}^2+\delta^2}$. Then
$$||f||_{L^4(S_{u,\ub})}\leq C\left(||f||_{L^2(S_{u,\ub})}+||\frac{<\phi,\nabla\phi>_{\gamma}}{\sqrt{|\phi|_{\gamma}^2+\delta^2}}||_{L^2(S_{u,\ub})}\right)\leq C\left(||f||_{L^2(S_{u,\ub})}+||\nabla\phi||_{L^2(S_{u,\ub})}\right).$$
The Proposition can be achieved by sending $\delta\to 0$.
\end{proof}
\begin{remark}
The dependence here on the initial data depends on the choice of coordinates on $S_{0,\ub}$. It is possible to instead work geometrically by considering the isoperimetric constant. We refer the reader to Section 5.2 in \cite{Chr} for more on this approach.
\end{remark}
The exact same proof gives control over the $L^6(S_{u,\ub})$ norm:
\begin{proposition}\label{L6}
There exists $\epsilon_0=\epsilon_0(\Delta_0,\mathcal G_0)$ such that as long as $\epsilon\leq \epsilon_0$, we have
$$||\phi||_{L^6(S_{u,\ub})}\leq C(\mathcal G_0)\sum_{i=0}^1||\nabla^i\phi||_{L^2(S_{u,\ub})}. $$
\end{proposition}
We can also prove the Sobolev Embedding Theorem for the $L^\infty$ norm: 
\begin{proposition}\label{Linfty}
There exists $\epsilon_0=\epsilon_0(\Delta_0,\mathcal G_0)$ such that as long as $\epsilon\leq \epsilon_0$, we have
$$||\phi||_{L^\infty(S_{u,\ub})}\leq C(\mathcal G_0)\left(||\phi||_{L^2(S_{u,\ub})}+||\nabla\phi||_{L^4(S_{u,\ub})}\right). $$
As a consequence,
$$||\phi||_{L^\infty(S_{u,\ub})}\leq C(\mathcal G_0)\sum_{i=0}^2||\nabla^i\phi||_{L^2(S_{u,\ub})}. $$
\end{proposition}
\begin{proof}
The first statement follows from coordinate considerations as in Proposition \ref{L4}. The second statement follows from applying the first and Proposition \ref{L4}.
\end{proof}
Besides the Sobolev Embedding Theorem on the 2-spheres, we also have a co-dimensional 1 trace formula that allows us to control the $L^4(S)$ norm by using the $L^2(H)$ or $L^2(\Hb)$ norm.
\begin{proposition}\label{trace}
Assume
$$\sup_{u,\ub}||\trch,\trchb||_{L^\infty(S_{u,\ub})}\leq 4\mathcal O_0.$$
Then
$$||\phi||_{L^4(S_{u,\ub})}\leq C(\mathcal O_0, \mathcal G_0)(||\phi||_{L^4(S_{u,0})}+||\phi||_{L^2(H_u)}^{\frac{1}{2}}||\nabla_4\phi||_{L^2(H_u)}^{\frac{1}{4}}(||\phi||_{L^2(H_u)}+||\nabla\phi||_{L^2(H_u)})^{\frac{1}{4}}),$$
$$||\phi||_{L^4(S_{u,\ub})}\leq 2(||\phi||_{L^4(S_{0,\ub})}+C(\mathcal G_0)||\phi||_{L^2(\Hb_{\ub})}^{\frac{1}{2}}||\nabla_3\phi||_{L^2(\Hb_{\ub})}^{\frac{1}{4}}(||\phi||_{L^2(\Hb_{\ub})}+||\nabla\phi||_{L^2(\Hb_{\ub})})^{\frac{1}{4}}),$$
\end{proposition}
\begin{proof}
The proof is standard (see for example \cite{KN}). We reproduce it here to emphasis that the constant is dependent only on the initial data. We will prove the first statement. The second statement follows analogously.
Using (\ref{L4transport}) and Proposition \ref{L6}, we have
\begin{equation*}
\begin{split}
 ||\phi||^4_{L^4(S_{u,\ub})}=&||\phi||^4_{L^4(S_{u,\ub'})}+\int_{\ub'}^{\ub}\int_{S_{u,\ub''}} 2\Omega|\phi|_\gamma^2\left(<\phi,\nabla_4\phi>_\gamma+ \frac{1}{2}\trch |\phi|^2_{\gamma}\right)d{\ub''}\\
\leq &||\phi||^4_{L^4(S_{u,\ub'})}+4||\phi||^3_{L^6(H)}||\nabla_4\phi||_{L^2(H)}+2\mathcal O_0\int_{0}^{\ub} ||\phi||_{L^4(S)}^4d{\ub''}\\
\leq &||\phi||^4_{L^4(S_{u,\ub'})}+2\mathcal O_0\int_{0}^{\ub} ||\phi||_{L^4(S_{u,\ub''})}^4d{\ub''}\\
&+C(\mathcal G_0)||\phi||_{L^4(H_u)}^2(||\phi||_{L^2(H_u)}+||\nabla\phi||_{L^2(H_u)})||\nabla_4\phi||_{L^2(H_u)}.
\end{split}
\end{equation*}
By Gronwall inequality, we have
\begin{equation*}
\begin{split}
 ||\phi||^4_{L^4(S_{u,\ub})}
\leq &C(\mathcal O_0, \mathcal G_0)(||\phi||^4_{L^4(S_{u,\ub'})}+||\phi||_{L^2(H_u)}^2(||\phi||_{L^2(H_u)}+||\nabla\phi||_{L^4(H_u)})||\nabla_4\phi||_{L^2(H_u)})\\
\end{split}
\end{equation*}
For the second statement, the proof follows analogously. However, since we are integrating in the $u$-direction, we can choose $\epsilon$ sufficiently small so that the constant is $2$.
\end{proof}

\subsection{Commutation Formulae}
We have the following for commutations:
\begin{proposition}
The commutator $[\nabla_4,\nabla]$ acting on an $(0,r)$ S-tensor is given by
\begin{equation*}
 \begin{split}
[\nabla_4,\nabla_B]\phi_{A_1...A_r}=&[D_4,D_B]\phi_{A_1...A_r}+(\nabla_B\log\Omega)\nabla_4\phi_{A_1...A_r}-(\gamma^{-1})^{CD}\chi_{BD}\nabla_C\phi_{A_1...A_r} \\
&-\sum_{i=1}^r (\gamma^{-1})^{CD}\chi_{BD}\etab_{A_i}\phi_{A_1...\hat{A_i}C...A_r}+\sum_{i=1}^r (\gamma^{-1})^{CD}\chi_{A_iB}\etab_{D}\phi_{A_1...\hat{A_i}C...A_r}.
 \end{split}
\end{equation*}
\end{proposition}
\begin{proof}
Since the formula is tensorial, it suffices to consider a basis ${e_1,e_2}$ on the 2-sphere that is orthonormal.
\begin{equation*}
 \begin{split}
\nabla_4\nabla_B\phi_{A_1...A_r}=D_4D_B\phi_{A_1...A_r}+\etab_B\nabla_4\phi_{A_1...A_r}-\sum_{i=1}^r \chi_{BC}\etab_{A_i}\phi_{A_1...\hat{A_i}C...A_r},
 \end{split}
\end{equation*}
where the notation in the last line means replacing the $i$-th slot by the index $C$.
\begin{equation*}
 \begin{split}
\nabla_B\nabla_4\phi_{A_1...A_r}=D_BD_4\phi_{A_1...A_r}-\zeta_B\nabla_4\phi_{A_1...A_r}+\chi_{BC}\nabla_C\phi_{A_1...A_r}-\sum_{i=1}^r \chi_{A_iB}\etab_{C}\phi_{A_1...\hat{A_i}C...A_r},
 \end{split}
\end{equation*}
Subtracting the later equation from the former, and using $\etab=\zeta+\nabla\log\Omega$, we can conclude the Proposition.
\end{proof}

By induction, we get the following schematic formula for multiple commutations:
\begin{proposition}
Suppose $\nabla_4\phi=F_0$. Let $\nabla_4\nabla^i\phi=F_i$.
Then
\begin{equation*}
\begin{split}
F_i\sim &\sum_{i_1+i_2+i_3=i}\nabla^{i_1}(\eta+\underline{\eta})^{i_2}\nabla^{i_3} F_0+\sum_{i_1+i_2+i_3+i_4=i}\nabla^{i_1}(\eta+\underline{\eta})^{i_2}\nabla^{i_3}\chi\nabla^{i_4} \phi\\
&+\sum_{i_1+i_2+i_3+i_4=i-1} \nabla^{i_1}(\eta+\underline{\eta})^{i_2}\nabla^{i_3}\beta\nabla^{i_4} \phi.
\end{split}
\end{equation*}
where by $\nabla^{i_1}(\eta+\underline{\eta})^{i_2}$ we mean the sum of all terms which is a product of $i_2$ factors, each factor being $\nabla^j (\eta+\underline{\eta})$ for some $j$ and that the sum of all $j$'s is $i_1$, i.e., $\nabla^{i_1}(\eta+\underline{\eta})^{i_2}=\displaystyle\sum_{j_1+...+j_{i_2}=i_1}\nabla^{j_1}(\eta+\underline{\eta})...\nabla^{j_{i_2}}(\eta+\underline{\eta})$. Similarly, suppose $\nabla_3\phi=G_{0}$. Let $\nabla_3\nabla^i\phi=G_{i}$.
Then
\begin{equation*}
\begin{split}
G_{i}\sim &\sum_{i_1+i_2+i_3=i}\nabla^{i_1}(\eta+\underline{\eta})^{i_2}\nabla^{i_3} G_{0}+\sum_{i_1+i_2+i_3+i_4=i}\nabla^{i_1}(\eta+\underline{\eta})^{i_2}\nabla^{i_3}\underline{\chi}\nabla^{i_4} \phi\\
&+\sum_{i_1+i_2+i_3+i_4=i-1} \nabla^{i_1}(\eta+\underline{\eta})^{i_2}\nabla^{i_3}\underline{\beta}\nabla^{i_4} \phi.
\end{split}
\end{equation*}

\end{proposition}
\begin{proof}
The proof is by induction. We will prove it for the first statement and the second one is analogous. It is obvious that the $i=0$ case is true. Assume that the statement is true for $i<i_0$.
\begin{equation*}
\begin{split}
F_{i_0}=&[\nabla_4,\nabla]\nabla^{i_0-1}\phi+\nabla F_{i_0-1}\\
\sim &\chi\nabla^{i_0}\phi+(\eta+\underline{\eta})\nabla_4\nabla^{i_0-1}\phi+\beta\nabla^{i_0-1}\phi+\chi(\eta+\underline{\eta})\nabla^{i_0-1}\phi\\
&+\sum_{i_1+i_2+i_3=i_0}\nabla^{i_1}(\eta+\underline{\eta})^{i_2}\nabla^{i_3} F_{0}\\
&+\sum_{i_1+i_2+i_3+i_4=i_0}\nabla^{i_1}(\eta+\underline{\eta})^{i_2}\nabla^{i_3}\chi\nabla^{i_4} \phi\\
&+\sum_{i_1+2i_2+i_3+i_4=i_0-1} \nabla^{i_1}(\eta+\underline{\eta})^{i_2}\nabla^{i_3}\beta\nabla^{i_4} \phi.
\end{split}
\end{equation*}
First notice that the first, third and fourth terms are acceptable. Then plug in the formula for $F_{i_0-1}=\nabla_4\nabla^{i_0-1}\phi$ and get the result.
\end{proof}
The following further simplified version is useful for our estimates in the next section:
\begin{proposition}
Suppose $\nabla_4\phi=F_0$. Let $\nabla_4\nabla^i\phi=F_i$.
Then
\begin{equation*}
\begin{split}
F_i\sim &\sum_{i_1+i_2+i_3=i}\nabla^{i_1}\psi^{i_2}\nabla^{i_3} F_0+\sum_{i_1+i_2+i_3+i_4=i}\nabla^{i_1}\psi^{i_2}\nabla^{i_3}\chi\nabla^{i_4} \phi.\\
\end{split}
\end{equation*}
Similarly, suppose $\nabla_3\phi=G_{0}$. Let $\nabla_3\nabla^i\phi=G_{i}$.
Then
\begin{equation*}
\begin{split}
G_{i}\sim &\sum_{i_1+i_2+i_3=i}\nabla^{i_1}\psi^{i_2}\nabla^{i_3} G_{0}+\sum_{i_1+i_2+i_3+i_4=i}\nabla^{i_1}\psi^{i_2}\nabla^{i_3}\underline{\chi}\nabla^{i_4} \phi.
\end{split}
\end{equation*}
\end{proposition}
\begin{proof}
We replace $\beta$ and $\betab$ using the Codazzi equations, which schematically looks like
$$\beta=\nabla\chi+\psi\chi,$$
$$\beta=\nabla\chib+\psi\chib.$$
\end{proof}

\subsection{The General Elliptic Estimates for the Hodge System}\label{elliptic}
We recall the definition of the divergence and curl of a symmetric covariant tensor of arbitrary rank:
$$(\div\phi)_{A_1...A_r}=\nabla^B\phi_{BA_1...A_r},$$
$$(\curl\phi)_{A_1...A_r}=\eps^{BC}\nabla_B\phi_{CA_1...A_r},$$
where $\eps$ is the volume form associated to the metric $\gamma$.
Recall also that the trace is defined to be
$$(tr\phi)_{A_1...A_{r-1}}=(\gamma^{-1})^{BC}\phi_{BCA_1...A_{r-1}}.$$
The following elliptic estimate is standard (See for example \cite{CK} or \cite{Chr}):
\begin{proposition}\label{ellipticbasic}
Let $\phi$ be a totally symmetric $r+1$ covariant tensorfield on a 2-sphere $(\mathbb S^2,\gamma)$ satisfying
$$\div\phi=f,\quad \curl\phi=g,\quad \mbox{tr}\phi=h.$$
Then
$$\int_{\mathbb S^2}(|\nabla\phi|^2+(r+1)K|\phi|^2)=\int_{\mathbb S^2}(|f|^2+|g|^2+rK|h|^2).$$
\end{proposition}
Given a totally symmetric $r+1$ covariant tensorfield $\phi$ on a 2-sphere $(\mathbb S^2, \gamma)$. We can define its totally symmetric derivative by
$$(\nabla\phi)^s_{CA_1...A_{r+1}}=\frac{1}{r+1}(\nabla_C\phi_{A_1...A_{r+1}}+\nabla_{A_1}\phi_{CA_2...A_{r+1}}+...+\nabla_{A_i}\phi_{CA_1...\hat{A_i}...A_{r+1}}+...+\nabla_{A_{r+1}}\phi_{CA_1...A_r}).$$
Then a direct computation would show (see \cite{Chr}, Lemma 7.2) that
\begin{proposition}\label{ellipder}
Let $\phi$ be a totally symmetric $r+1$ covariant tensorfield on a 2-sphere $(\mathbb S^2,\gamma)$ satisfying
$$\div\phi=f,\quad \curl\phi=g,\quad tr\phi=h.$$
Then $\phi'=(\nabla\phi)^s$ satisfies
$$\div\phi'=f',\quad \curl\phi'=g',\quad tr\phi'=h',$$
where
$$f'=(\nabla f)^s-\frac{1}{r+2}(^*\nabla g)^s+(r+1)\phi-\frac{2K}{r+1}(\gamma\otimes^s h),$$
$$g'=\frac{r+1}{r+2}(\nabla g)^s+(r+1)K(^*\phi)^s,$$
$$h'=\frac{2}{r+2}f+\frac{r}{r+2}(\nabla h)^s.$$
\end{proposition}
Inducting using the previous two Propositions, we can estimate an arbitrary number of derivatives of tensorfields.
\begin{proposition}\label{ellipticthm}
Let $\phi$ be a totally symmetric $r+1$ covariant tensorfield on a 2-sphere $(\mathbb S^2,\gamma)$ satisfying
$$\div\phi=f,\quad \curl\phi=g,\quad \mbox{tr}\phi=h.$$
Suppose also that
$$\sum_{i=0}^2||\nabla^i K||_{L^2(S)}\leq \infty.$$
Then for $i\leq 3$,
$$||\nabla^{i}\phi||_{L^2(S)}\leq C(\sum_{i=0}^2||\nabla^i K||_{L^2(S)},\mathcal G_0)\sum_{j=0}^{i-1}(||\nabla^{j}f||_{L^2(S)}+||\nabla^{j}g||_{L^2(S)}+||\nabla^{j}h||_{L^2(S)}+||\phi||_{L^2(S)})$$
\end{proposition}
\begin{proof}
In this proof, let $\phi^{(n)}$ denote the symmetrized n-th covariant derivative of $\phi$. This is defined inductively:
$$\phi'=(\nabla\phi)^s,$$
and
$$\phi^{(n+1)}=(\phi^{(n)})'.$$ 
Then $\phi^{(n)}$ satisfies the following generalized Hodge system:

We would also like to show that we can inductively recover all the derivatives from the symmetrized derivatives. It is instructive to first look at the case of 1 derivative. 
\begin{equation*}
\begin{split}
\nabla_{B}\phi_{A_1...A_{r+1}}=&\phi'_{BA_1...A_{r+1}}+\frac{1}{r+2}((r+1)\nabla_B\phi_{A_1...A_{r+1}}-\nabla_{A_1}\phi_{BA_2...A_{r+1}}-\nabla_{A_{r+1}}\phi_{BA_1...A_r}) \\
=&\phi'_{BA_1...A_{r+1}}+\frac{1}{r+2}(\eps_{BA_1}g_{A_2...A_{r+1}}+\eps_{BA_2}g_{A_1\hat{A_2}...A_{r+1}}+...+\eps_{BA_{r+1}}g_{A_2...A_{r}}).
\end{split}
\end{equation*}
To retrieve higher order derivatives, we claim that the following holds
\begin{equation*}
\begin{split}
\nabla_{B_1}...\nabla_{B_n}\phi_{A_1...A_{r+1}}-\phi^{(n)}_{B_1...B_nA_1...A_{r+1}}\sim f^{(n-1)}+g^{(n-1)}+\sum_{i+j\leq n-1}\phi^{(j)}\nabla^{i}K
\end{split}
\end{equation*}
This can be proved inductively. Finally, using Proposition \ref{ellipder}, we can derive inductively a div-curl system for $\phi^{(n)}$, which would allow us to conclude using Proposition \ref{ellipticbasic}, as long as we can control $K$. The terms that we need to control are
$$\sum_{i+j\leq 2}||\phi^{(j)}\nabla^iK||_{L^2(S)} \leq C(||\phi||_{L^\infty(S)}||\nabla^2K||_{L^2(S)}+||\nabla\phi||_{L^4(S)}||\nabla K||_{L^4(S)}+||\nabla^2\phi||_{L^2(S)}||K||_{L^2(S)}).$$
By Sobolev Embedding in Propositions \ref{L4} and \ref{Linfty}, we have
$$||\phi||_{L^\infty(S)}+||\nabla\phi||_{L^4(S)}\leq ||\nabla^2\phi||_{L^2(S)}+||\phi||_{L^2(S)}.$$
We also need to control terms involving $K$ and $f, g, h$. Using Sobolev Embedding in Propositions \ref{L4} and \ref{Linfty}, we have
$$||K\nabla^2 h||_{L^2(S)}\leq ||K||_{L^\infty(S)}||\nabla^2 h||_{L^2(S)}\leq C(\mathcal G_0)\sum_{i=0}^2||\nabla^iK||_{L^2(S)}||\nabla^2 h||_{L^2(S)},$$
$$||K\nabla (f,g)||_{L^2(S)}\leq ||K||_{L^\infty(S)}||\nabla(f,g)||_{L^2(S)}\leq C(\mathcal G_0)\sum_{i=0}^2||\nabla^iK||_{L^2(S)}||\nabla (f,g)||_{L^2(S)},$$
$$||K\nabla h||_{L^2(S)}\leq ||K||_{L^\infty(S)}||\nabla h||_{L^2(S)}\leq C(\mathcal G_0)\sum_{i=0}^2||\nabla^iK||_{L^2(S)}||\nabla h||_{L^2(S)},$$
$$||\nabla K h||_{L^2(S)}\leq ||\nabla K||_{L^4(S)}||h||_{L^4(S)}\leq C(\mathcal G_0)\sum_{i=0}^2||\nabla^iK||_{L^2(S)}(||\nabla h||_{L^2(S)}+||\nabla^2 h||_{L^2(S)}).$$
\end{proof}
In the following, we will only apply the above Proposition for $\phi$ a symmetric traceless 2-tensor. For such tensors, it suffices to know its divergence:
\begin{proposition}\label{ellipticchih}
Suppose $\phi$ is a symmetric traceless 2-tensor satisfying
$$\div\phi=f.$$
Suppose moreover that $$\sum_{i=0}^2||\nabla^i K||_{L^2(S)}\leq \infty.$$
Then, for $i\leq 3$,
$$||\nabla^{i}\phi||_{L^2(S)}\leq C(\sum_{i=0}^2||\nabla^i K||_{L^2(S)},\mathcal G_0)\sum_{j=0}^{2}(||\nabla^{j}f||_{L^2(S)}+||\phi||_{L^2(S)}).$$
\end{proposition}
\begin{proof}
In view of the previous Proposition, this Proposition follows from
$$\curl\phi=^*f.$$
This is a direct computation using that fact that $\phi$ is both symmetric and traceless.
\end{proof}

\section{Estimates}\label{estimates}

Given $C^\infty$ characteristic initial data, Rendall's Theorem guarantees a smooth solution in a neighborhood of the intersection of the two null hypersurfaces. We will prove a priori estimates for this solution up to three derivatives of the curvature in $L^2$. This would then be sufficient to run a standard persistence of regularity argument to show that the solution remains smooth. We can then use a ``last slice'' argument to show that the solution indeed exist in the full region in which we can prove estimates.

We define the initial data norms as follows: 
Let $$\mathcal O_0=\sup_{S\subset H_0,S\subset \Hb_0}\sup_{\psi\in\{\chih,\trch,\omega,\eta,\etab,\chibh,\trchb,\omegab\}}\max\{1,\sum_{i=0}^{3}||\nabla^i\psi||_{L^2(S)},\sum_{i=0}^{2}||\nabla^i\psi||_{L^4(S)},\sum_{i=0}^1||\nabla^i\psi||_{L^\infty(S)}\}.$$
\begin{equation*}
\begin{split}
\mathcal R_0=&
\sum_{i=0}^{3}\left(\sup_{\Psi\in\{\alpha,\beta,\rho,\sigma,\betab\}}||\nabla^i\Psi||_{L^2(H_0)}+\sup_{\Psi\in\{\beta,\rho,\sigma,\betab,\alphab\}}||\nabla^i\Psi||_{L^2(\Hb_0)}\right)\\
&+\sup_{S\subset H_0,S\subset \Hb_0}\sup_{\Psi\in\{\alpha,\beta,\rho,\sigma,\betab,\alphab\}}\max\{1,\sum_{i=0}^{2}||\nabla^i\Psi||_{L^2(S)},\sum_{i=0}^1||\nabla^i\Psi||_{L^4(S)}\}.
\end{split}
\end{equation*}

Notice that the definition of $\mathcal R_0$ includes terms that are integrated on the hypersurfaces as well as terms that are integrated on 2-spheres. These numbers are fixed by the initial data. They are assumed to be finite. Whenever we would like to refer to a constant that depends only on the initial data, we will use $C(\mathcal O_0, \mathcal R_0)$ (or $C(\mathcal O_0)$ etc.). 

Let $$\mathcal R= \sum_{i=0}^{3}\sup_u\sup_{\Psi\in\{\alpha,\beta,\rho,\sigma,\betab\}}||\nabla^i\Psi||_{L^2(H_u)}+\sup_{\ub}\sup_{\Psi\in\{\beta,\rho,\sigma,\betab,\alphab\}}||\nabla^i\Psi||_{L^2(\Hb_{\ub})},$$
$$\mathcal R(S)=\sum_{i=0}^2 \sup_{u,\ub}||\nabla^i(\alpha,\beta,\rho,\sigma,\betab)||_{L^4(S_{u,\ub})}.$$

Our main strategy will be as follows:
We will prove all the estimates in two steps. In the first step, we will assume the boundedness of $\mathcal R$ and prove that $\epsilon$ can be chosen so that the Ricci coefficients can be controlled by $\mathcal R(S)$ and the initial data. This can be achieved by considering the null structure equations (\ref{null.str1}) and (\ref{null.str2}) as transport equations for the Ricci coefficients, with the null curvature components as source. We then use Propositions \ref{transport} and \ref{transportinfty} to derive the necessary estimates. As we have mentioned in the introduction, the main observation in deriving these estimates is that either we have a $\nabla_3$ equation, for which we have a smallness constant $\epsilon$ and can derive the required estimates; or that whenever we have a $\nabla_4$ equation, it must be the case that in the nonlinear term $\Gamma\cdot\Gamma$, at least one factor has already been estimated either by a $\nabla_3$ equation or by a $\nabla_4$ equation that we have considered first. This would allow the $\nabla_4$ equations to be considered as essentially \emph{linear} equations and thus we can derive the necessary estimates by Gronwall inequality. In order to make this strategy work, we need to track which of the terms can be controlled by the size of the initial data alone.

The second step is the energy estimates for the curvature components, i.e., the estimates for $\mathcal R$. We will derive, by simple integration by parts, the energy estimates that shows that $\mathcal R$ can be controlled by $\mathcal R_0$ and nonlinear terms. The nonlinear terms would have the form (neglecting derivatives) $\psi\Psi\Psi$ and the energy estimates schematically look like:
$$\mathcal R^2 \leq \mathcal R_0+\int_{D_{u,\ub}} \psi\Psi\Psi.$$
Notice that using the norm $\mathcal R$, all curvature components except $\alphab$ can be controlled on $H_u$. Thus, whenever these components appear in the nonlinear term, we can control them on $H_u$ and integrate in $u$, thus gaining a smallness constant $\epsilon$. The main difficulty comes from the terms of the form $\psi\alphab^2$. For these terms, we show that $\psi$ can in fact be controlled (in the first step) by the initial data alone. Thus the nonlinear term is essentially \emph{linear} and we can close the estimates by Gronwall inequality.
It is precisely for this reason that we separated the $\mathcal R$ and the $\mathcal R(S)$ norms. We would like to show that the $\mathcal R(S)$ norms can be controlled by initial data alone, i.e., independent of $\mathcal R$ and thus allow us to control some norms for the Ricci coefficients independent of $\mathcal R$. We can then use this to close the energy estimates.

We now perform step one of the proof. We will first show the $L^\infty$ bounds for the Ricci coefficients.

\begin{proposition}\label{RicciLinfty}
Assume
$$\mathcal R<\infty, \quad\mathcal R(S)< \infty,\quad \sup_{u,\ub}||\nabla^3\eta||_{L^2(S_{u,\ub})}< \infty.$$
Then there exists $\epsilon_0=\epsilon_0(\mathcal O_0, \mathcal R_0,\displaystyle\sup_{u,\ub}||\nabla^3\eta||_{L^2(S_{u,\ub})},\mathcal R, \mathcal G_0)$ such that for $\epsilon\leq \epsilon_0$, we have
$$\sup_{u,\ub}||\chih,\trch,\chibh,\trchb,\etab,\omega||_{L^\infty(S_{u,\ub})}\leq 3\mathcal O_0, $$
$$\sup_{u,\ub}||\eta,\omegab||_{L^\infty(S_{u,\ub})}\leq C(\mathcal O_0,\mathcal R(S),\mathcal G_0).$$
\end{proposition}
\begin{proof}
It suffices to prove the estimates under the bootstrap assumption:
\begin{equation}\label{RicciLinftyBA}
||\chih,\trch,\chibh,\trchb,\etab,\omega||_{L^\infty(S)}\leq 4\mathcal O_0 .
\end{equation}
This in particular allows us to use Proposition \ref{transportinfty}.
We first estimate the $L^\infty$ norm of $\eta$. Notice that $\omegab$ and $\eta\eta$ do not appear in the null structure equation for $\nabla_4\eta$. Notice also that in this null structure equation, the curvature term does not contain $\alphab$. Thus, using the Sobolev Embedding in Proposition \ref{Linfty}, we can estimate the curvature term using $\mathcal R(S)$. Thus,
$$||\eta||_{L^\infty(S_{u,\ub})}\leq \mathcal O_0+CI\mathcal R(S)+C(\mathcal O_0)(1+\int_0^{\ub} ||\eta||_{L^\infty(S_{u,\ub'})}d\ub').$$
By Gronwall's inequality,
$$||\eta||_{L^\infty(S_{u,\ub})}\leq C(\mathcal O_0,\mathcal R(S), \mathcal G_0)\exp(C(\mathcal O_0)\ub).$$
Since $\ub\leq I$, we clearly have
$$\sup_{u,\ub}||\eta||_{L^\infty(S_{u,\ub})}\leq C(\mathcal O_0,\mathcal R(S), \mathcal G_0).$$
We then move to the estimate for the $L^\infty$ norm of $\omegab$. Here, the key is to observe that $\omegab^2$ does not appear in the equation for $\nabla_4\omegab$. Moreover, as for the equation for $\nabla_4\eta$, the curvature term can be controlled by $\mathcal R(S)$. Thus,
$$||\omegab||_{L^\infty(S_{u,\ub})}\leq \mathcal O_0+CI \mathcal R(S)+C(\mathcal O_0,\mathcal R(S), \mathcal G_0)+C(\mathcal O_0)\int_0^{\ub} ||\omegab||_{L^\infty(S_{u,\ub'})}d\ub'.$$
As for the estimates for $\eta$, we use Gronwall's inequality to get
$$||\omegab||_{L^\infty(S_{u,\ub})}\leq C(\mathcal O_0,\mathcal R(S), \mathcal G_0)\exp(C(\mathcal O_0)\ub).$$
As before, since $\ub\leq I$, we clearly have
$$\sup_{u,\ub}||\omegab||_{L^\infty(S_{u,\ub})}\leq C(\mathcal O_0,\mathcal R(S), \mathcal G_0).$$
We now have to close the bootstrap assumptions for $\chih,\trch,\chibh,\trchb,\etab,\omega$. We have
\begin{equation*}
\begin{split}
\nabla_3(\chih,\trch)=\Psi+\nabla\eta+\psi\psi,
\end{split}
\end{equation*}
and
\begin{equation*}
\begin{split}
\nabla_3(\chibh,\trchb,\etab,\omega)=\Psi+\psi\psi.
\end{split}
\end{equation*}
Using these it follows from Proposition \ref{transport} that 
$$||\chih,\trch||_{L^\infty(S)}\leq 2\mathcal O_0+C\epsilon\sup_{u,\ub}||\nabla^3\eta||_{L^2(S_{u,\ub})}+\epsilon C(\mathcal O_0,\mathcal R, \mathcal G_0).$$
$$||\chibh,\trchb,\etab,\omega||_{L^\infty(S)}\leq 2\mathcal O_0+C\epsilon^{\frac{1}{2}}\mathcal R+\epsilon C(\mathcal O_0,\mathcal R, \mathcal G_0).$$
By choosing $\epsilon$ sufficiently small depending on $\mathcal O_0, \mathcal R(S), \displaystyle\sup_{u,\ub}||\nabla^3\eta||_{L^2(S_{u,\ub})}, I$, we have
$$\sup_{u,\ub}||(\chih,\trch,\chibh,\trchb,\etab,\omega)||_{L^\infty(S_{u,\ub})}\leq 3\mathcal O_0,$$
hence improving the bootstrap assumption (\ref{RicciLinftyBA}).
\end{proof}

We will estimate the $L^4$ norms using a similar strategy. 
\begin{proposition}\label{RicciL4}
With the same assumptions as in the last proposition, we have some $$\epsilon_0=\epsilon_0(\mathcal O_0,\mathcal R(S),\sup_{u,\ub}||\nabla^3\eta||_{L^2(S_{u,\ub})},\mathcal R, \mathcal G_0)$$ such that for $\epsilon\leq \epsilon_0$, we have
$$||\nabla(\chih,\trch,\chibh,\trchb,\etab,\omega)||_{L^4(S)}\leq 3\mathcal O_0, $$
$$||\nabla(\eta,\omegab)||_{L^4(S)}\leq C(\mathcal O_0,\mathcal R(S), \mathcal G_0).$$
\end{proposition}
\begin{proof}
We introduce the bootstrap assumptions:
\begin{equation}\label{RicciL4BA}
||\nabla(\chih,\trch,\chibh,\trchb,\etab,\omega)||_{L^4(S)}\leq 4\mathcal O_0.
\end{equation}

In order to estimate the $L^4$ norm of the Ricci coefficient, we will use Proposition \ref{transport}. It is applicable since the previous Proposition implies the necessary bounds on $||\trch,\trchb||_{L^\infty(S)}$. We first estimate the $L^4$ norm of $\nabla\eta$. Notice that $\nabla\omegab$, $(\nabla\eta)^2$ do not appear in the null structure equations. The term $\eta\nabla\eta$ appears, but we can estimate it by
$$||\eta\nabla\eta||_{L^4(S_{u,\ub})}\leq ||\eta||_{L^\infty(S_{u,\ub})}||\nabla\eta||_{L^4(S_{u,\ub})}$$
and use the estimate for $||\eta||_{L^\infty(S_{u,\ub})}$ from the previous Proposition. Moreover, notice that $\alphab$ does not appear in the equation and by Sobolev Embedding in Proposition \ref{L4}, we can control the curvature term by $\mathcal R(S)$. Putting all these observations together and using Proposition \ref{transport} and \ref{RicciLinfty} and (\ref{RicciL4BA}), we have
$$||\nabla\eta||_{L^4(S_{u,\ub})}\leq C(\mathcal O_0,\mathcal R(S), \mathcal G_0)+C(\mathcal O_0,\mathcal R(S), \mathcal G_0)\int_0^{\ub} ||\nabla\eta||_{L^4(S_{u,\ub'})}d\ub'.$$
By Gronwall's inequality,
$$||\nabla\eta||_{L^4(S_{u,\ub})}\leq C(\mathcal O_0,\mathcal R(S), \mathcal G_0)\exp(C(\mathcal O_0,\mathcal R(S), \mathcal G_0)\ub).$$
Therefore,
$$||\nabla\eta||_{L^4(S_{u,\ub})}\leq C(\mathcal O_0,\mathcal R(S), \mathcal G_0).$$
We then move to the estimate for the $L^4$ norm of $\nabla\omegab$. Here, the key is to observe that $(\nabla\omegab)^2$ does not appear in the equation for $\nabla_4\nabla\omegab$ and that $\alphab$ does not appear as the curvature term. Using Proposition \ref{transport} and \ref{RicciLinfty} and (\ref{RicciL4BA}), we have,
$$||\nabla\omegab||_{L^4(S_{u,\ub})}\leq C(\mathcal O_0,\mathcal R(S), \mathcal G_0)+C(\mathcal O_0,\mathcal R(S), \mathcal G_0)\int_0^{\ub} ||\nabla\omega||_{L^4(S_{u,\ub'})}d\ub'.$$
As for the estimates for $\nabla\eta$, we use Gronwall's inequality to get
$$||\nabla\omegab||_{L^4(S_{u,\ub})}\leq C(\mathcal O_0,\mathcal R(S), \mathcal G_0)\exp(C(\mathcal O_0,\mathcal R(S), \mathcal G_0)\ub).$$
Therefore,
$$||\nabla\omegab||_{L^4(S_{u,\ub})}\leq C(\mathcal O_0,\mathcal R(S), \mathcal G_0).$$
Moreover, as in the proof of Proposition \ref{RicciLinfty},
\begin{equation*}
\begin{split}
&||\nabla(\chih,\trch,\chibh,\trchb,\etab,\omega)||_{L^4(S_{u,\ub})}\\
\leq & 2\mathcal O_0+C(\epsilon^{\frac{1}{2}}\mathcal R+\epsilon\sup_{u,\ub}||\nabla^3\eta||_{L^2(S_{u,\ub})})+\epsilon C(\mathcal O_0,\mathcal R(S), \mathcal G_0)\sup_{\ub'\in [0,\ub]}||\nabla(\chih,\trch,\chibh,\trchb,\etab,\omega)||_{L^4(S_{u,\ub'})}.
\end{split}
\end{equation*}
Hence, by choosing $\epsilon$ sufficiently small depending on $\mathcal O_0, \mathcal R(S), \displaystyle\sup_{u,\ub}||\nabla^3\eta||_{L^2(S_{u,\ub})},\mathcal R, I$, we have
$$\sup_{u,\ub}||\nabla(\chih,\trch,\chibh,\trchb,\etab,\omega)||_{L^4(S_{u,\ub})}\leq 3\mathcal O_0,$$
which improves the bootstrap assumption and gives the desired estimates.
\end{proof}
We now move on to the $L^2$ norm of the Ricci coefficients:
\begin{proposition}\label{RicciL2}
With the same assumptions as in the Proposition \ref{RicciLinfty}, we have some $$\epsilon_0=\epsilon_0(\mathcal O_0,\mathcal R(S),\displaystyle\sup_{u,\ub}||\nabla^3\eta||_{L^2(S_{u,\ub})},\mathcal R, \mathcal G_0)$$ such that for $\epsilon\leq \epsilon_0$, we have
$$\sup_{u,\ub}||\nabla^2(\chih,\trch,\chibh,\trchb,\etab,\omega)||_{L^2(S_{u,\ub})}\leq 3\mathcal O_0, $$
$$\sup_{u,\ub}||\nabla^2(\eta,\omegab)||_{L^2(S_{u,\ub})}\leq C(\mathcal O_0,\mathcal R(S), \mathcal G_0)$$
\end{proposition}
\begin{proof}
We make the bootstrap assumption:
$$\sup_{u,\ub}||\nabla(\chih,\trch,\chibh,\trchb,\etab,\omega)||_{L^2(S_{u,\ub})}\leq 4\mathcal O_0.$$
As before, we first estimate $\nabla^2\eta$ and $\nabla^2\omegab$. Using the $L^\infty$ bounds for $\psi$ and $L^4$ bounds for $\nabla\psi$, we have
$$||\nabla^2\eta||_{L^2(S_{u,\ub})}\leq C(\mathcal O_0,\mathcal R(S), \mathcal G_0)+C(\mathcal O_0,\mathcal R(S), \mathcal G_0)\int_0^{\ub} ||\nabla^2\eta||_{L^2(S_{u,\ub'})}d\ub'.$$
By Gronwall's inequality,
$$||\nabla^2\eta||_{L^2(S_{u,\ub})}\leq C(\mathcal O_0,\mathcal R(S), \mathcal G_0).$$
We then move to the estimate for the $L^\infty$ norm of $\omegab$. Here, the key is to observe that $\omegab^2$ and $\eta\omegab$ do not appear in the equation for $\nabla_4\omegab$.
$$||\nabla^2\omegab||_{L^2(S_{u,\ub})}\leq C(\mathcal O_0,\mathcal R(S), \mathcal G_0)+C(\mathcal O_0,\mathcal R(S), \mathcal G_0)\int_0^{\ub} ||\nabla^2\omega||_{L^2(S_{u,\ub'})}d\ub'.$$
As for the estimates for $\nabla\eta$, we use Gronwall's inequality to get
$$||\nabla^2\omegab||_{L^2(S_{u,\ub})}\leq C(\mathcal O_0,\mathcal R(S), \mathcal G_0).$$
Moreover,
$$||\nabla(\chih,\trch,\chibh,\trchb,\etab,\omega)||_{L^2(S_{u,\ub})}\leq 2\mathcal O_0+C\epsilon^{\frac{1}{2}}\mathcal R+\epsilon C(\mathcal O_0,\mathcal R(S), \mathcal G_0)\sup_{\ub'\in [0,\ub]}||\nabla(\chih,\trch,\chibh,\trchb,\etab,\omega)||_{L^2(S_{u,\ub'})}.$$
Hence, by choosing $\epsilon$ sufficiently small depending on $\mathcal O_0,\mathcal R(S),\displaystyle\sup_{u,\ub}||\nabla^3\eta||_{L^2(S_{u,\ub})},\mathcal R,I$, we have
$$||\nabla(\chih,\trch,\chibh,\trchb,\etab,\omega)||_{L^2(S)}\leq 3\mathcal O_0,$$
which improves the bootstrap assumption and gives the desired estimates.
\end{proof}
We have now proved that the estimates for the Ricci coefficients up to two derivatives in $L^2(S)$ can be estimated by a constant depending only on the initial data and $\mathcal R(S)$. This is enough for us to show that the $\mathcal R(S)$ norms can be controlled by the initial data alone. This in turn implies that the Ricci coefficients can be controlled by initial data alone.
\begin{proposition}\label{RS}
Assume
$$\mathcal R< \infty.$$
Then there exists $\epsilon_0=\epsilon_0(\mathcal O_0,\mathcal R_0, \mathcal R)$ such that for $\epsilon\leq \epsilon_0$, we have
$$\mathcal R(S)\leq C(\mathcal R_0).$$
\end{proposition}
\begin{proof}
We first prove the weaker statement
$$\sum_{i=0}^2 ||\nabla^i(\alpha,\beta,\rho,\sigma,\betab)||_{L^2(S_{u,\ub})}\leq C(\mathcal R_0).$$
This can be achieved by considering the $\nabla_3$ null Bianchi equations as transport equations and estimate using Proposition \ref{transport}. 
\begin{equation*}
\begin{split}
&\sum_{i=0}^2 ||\nabla^i(\alpha,\beta,\rho,\sigma,\betab)||_{L^2(S_{u,\ub})}\\
\leq &2\mathcal R_0+\epsilon^{\frac{1}{2}}\sum_{\Psi\in\{\beta,\rho,\sigma,\betab,\alphab\}}\sum_{i=0}^3 ||\nabla^i\Psi||_{L^2(\Hb_{\ub})}+\sum_{i_1+i_2+i_3+i_4\leq 2}\sum_{\Psi\in\{\alpha,\beta,\rho,\sigma,\betab,\alphab\}}||\nabla^{i_1}\psi^{i_2}\nabla^{i_3}\psi\nabla^{i_4}\Psi||_{L^1_{u}L^2(S)}
\end{split}
\end{equation*}
By Proposition \ref{RicciL2}, 
$$\sup_{u,\ub}\sum_{i=0}^2||\nabla^i\psi||_{L^2(S_{u,\ub})}\leq C(\mathcal O_0,\mathcal R(S), \mathcal G_0).$$
Therefore,
$$\sum_{i_1+i_2+i_3+i_4\leq 2}||\nabla^{i_1}\psi^{i_2}\nabla^{i_3}\psi\nabla^{i_4}\Psi||_{L^1_{u}L^2(S)}\leq \epsilon^{\frac{1}{2}}C(\mathcal O_0, \mathcal R(S), \mathcal G_0)\mathcal R+C(\mathcal O_0, \mathcal R(S), \mathcal G_0)\sum_{i=0}^2\int_0^{u} ||\nabla^i\alpha||_{L^2(S_{u',\ub})} du'.$$
Hence, by choosing $\epsilon$ sufficiently small, we have 
$$\sum_{i=0}^2 ||\nabla^i(\alpha,\beta,\rho,\sigma,\betab)||_{L^2(S_{u,\ub})}\leq 3\mathcal R_0+C(\mathcal O_0, \mathcal R(S), \mathcal G_0)\sum_{i=0}^2\int_0^{u} ||\nabla^i\alpha||_{L^2(S_{u',\ub})} du'.$$
By Gronwall's inequality and that $u\leq \epsilon$, we thus have
$$\sum_{i=0}^2 ||\nabla^i(\alpha,\beta,\rho,\sigma,\betab)||_{L^2(S_{u,\ub})}\leq 4\mathcal R_0.$$
We now estimate the $L^4(S)$ norms of $\nabla^i(\alpha,\beta,\rho,\sigma,\betab)$ using the codimension 1 trace formula in Proposition \ref{trace}. Take $\phi=\nabla^i(\alpha,\beta,\rho,\sigma,\betab)$ for $i\leq 2$ and substitute into
$$||\phi||_{L^4(S_{u,\ub})}\leq 2(||\phi||_{L^4(S_{0,\ub})}+C(\mathcal G_0)||\phi||_{L^2(\Hb_{\ub})}^{\frac{1}{2}}||\nabla_3\phi||_{L^2(\Hb_{\ub})}^{\frac{1}{4}}(||\phi||_{L^2(\Hb_{\ub})}+||\nabla\phi||_{L^2(\Hb_{\ub})})^{\frac{1}{4}}).$$
By definition of the $\mathcal R$ and $\mathcal R(S)$ norms, we have
$$\sum_{i=0}^2||\nabla(\nabla^i(\alpha,\beta,\rho,\sigma,\betab))||_{L^2(\Hb_{\ub})}\leq C(\mathcal O_0,\mathcal R(S),\mathcal R, \mathcal G_0).$$
and, using the null Bianchi equations, we also have
\begin{equation*}
\begin{split}
&\sum_{i=0}^2||\nabla_3(\nabla^i(\alpha,\beta,\rho,\sigma,\betab))||_{L^2(\Hb_{\ub})} \\
\leq & \sum_{i=0}^3\sum_{\Psi\in\{\beta,\rho,\sigma,\betab,\alphab\}}||\nabla^i\Psi||_{L^2(\Hb_{\ub})}+\sum_{i_1+i_2+i_3+i_4\leq 2}\sum_{\Psi\in\{\alpha,\beta,\rho,\sigma,\betab,\alphab\}}||\nabla^{i_1}\psi^{i_2}\nabla^{i_3}\psi\nabla^{i_4}\Psi||_{L^2(\Hb_{\ub})}.
\end{split}
\end{equation*}
The first term is clearly $\leq C\mathcal R$. Using Sobolev Embedding in Propositions \ref{L4} and \ref{Linfty}, as well as the estimates for the Ricci coefficients from Propositions \ref{RicciLinfty}, \ref{RicciL4} and \ref{RicciL2}, we have
$$\sum_{i_1+i_2+i_3+i_4\leq 2}\sum_{\Psi\in\{\alpha,\beta,\rho,\sigma,\betab,\alphab\}}||\nabla^{i_1}\psi^{i_2}\nabla^{i_3}\psi\nabla^{i_4}\Psi||_{L^2(\Hb_{\ub})}\leq C(\mathcal O_0,\mathcal R(S),\mathcal G_0)\sum_{i=0}^2||\nabla^i\Psi||_{L^2(\Hb_{\ub})}.$$
Notice that all components except $\alpha$ can be controlled in $L^2(\Hb_{\ub})$, we have
$$\sum_{i=0}^2||\nabla_3(\nabla^i(\alpha,\beta,\rho,\sigma,\betab))||_{L^2(\Hb_{\ub})}\leq C\mathcal R+C(\mathcal O_0,\mathcal R(S),\mathcal R_0,\mathcal G_0)(\mathcal R+\epsilon^{\frac 12}\sup_{u,\ub}||\alpha||_{L^2(S_{u,\ub})}.$$
The smallness parameter that we need for the estimates comes in through the $L^2(S)$ estimates that we have just derived:
$$\sum_{i=0}^2||\nabla^i(\alpha,\beta,\rho,\sigma,\betab)||_{L^2(\Hb_{\ub})}\leq \epsilon^{\frac{1}{2}}\sup_{u,\ub}\sum_{i=0}^2||\nabla^i(\alpha,\beta,\rho,\sigma,\betab)||_{L^2(S_{u,\ub})}\leq \epsilon^{\frac{1}{2}}C(\mathcal O_0,\mathcal R(S),\mathcal R, \mathcal G_0).$$
Therefore,
$$\sum_{i=0}^2||\nabla^i(\alpha,\beta,\rho,\sigma,\betab)||_{L^4(S_{u,\ub})}\leq 2(\sum_{i=0}^2||\nabla^i(\alpha,\beta,\rho,\sigma,\betab)||_{L^4(S_{0,\ub})}+\epsilon^{\frac{1}{4}}C(\mathcal O_0,\mathcal R(S),\mathcal R, \mathcal G_0)).$$
The necessary estimate follows by choosing $\epsilon$ sufficiently small. 
\end{proof}
This implies that in Propositions \ref{RicciLinfty}, \ref{RicciL4} and \ref{RicciL2},the bounds can be in terms of initial data alone. In other words,
\begin{proposition}\label{Ricci}
Assume
$$\mathcal R<\infty,\quad \sup_{u,\ub}||\nabla^3\eta||_{L^2(S_{u,\ub})}< \infty.$$
Then there exists $\epsilon_0=\epsilon_0(\mathcal O_0, \mathcal R(S),\displaystyle\sup_{u,\ub}||\nabla^3\eta||_{L^2(S_{u,\ub})},\mathcal R, \mathcal G_0)$ such that for $\epsilon\leq \epsilon_0$, we have
$$||\psi||_{L^\infty(S_{u,\ub})}\leq C(\mathcal O_0,\mathcal R_0,\mathcal G_0), $$
$$\sum_{i=0}^1||\nabla^i\psi||_{L^4(S_{u,\ub})}\leq C(\mathcal O_0,\mathcal R_0,\mathcal G_0),$$
$$\sum_{i=0}^2||\nabla^i\psi||_{L^2(S_{u,\ub})}\leq C(\mathcal O_0,\mathcal R_0,\mathcal G_0).$$
\end{proposition}
These estimates allow us also to derive estimates for the Gauss curvature $K$ on the spheres.
\begin{proposition}\label{Kest}
Assume
$$\mathcal R< \infty.$$
Then there exists $\epsilon_0=\epsilon_0(\mathcal O_0,\mathcal R_0, \mathcal R)$ such that for $\epsilon\leq \epsilon_0$, we have
$$\sum_{i=0}^2||\nabla^iK||_{L^2(S)}\leq C(\mathcal O_0, \mathcal R_0, \mathcal G_0).$$
\end{proposition}
\begin{proof}
We estimate $\nabla^iK$ in $L^4(S)$ using the Gauss equation
$$K=-\rho+\frac{1}{2}\chih\cdot\chibh-\frac{1}{4}\trch\cdot \trchb.$$
Proposition \ref{RS}, \ref{area} and Holder's inequality give
$$\sum_{i=0}^2||\nabla^i\rho||_{L^2(S_{u,\ub})}\leq  C(\mathcal O_0, \mathcal R_0, \mathcal G_0).$$
The $\mathcal R(S)$ estimates in Proposition \ref{RS}, together with Proposition \ref{RicciL2}, give
$$\sum_{i=0}^2||\nabla^i(\trch,\trchb,\chih,\chibh)||_{L^2(S_{u,\ub})}\leq C(\mathcal O_0, \mathcal R_0, \mathcal G_0).$$
The $\mathcal R(S)$ estimates in Proposition \ref{RS}, together with Proposition \ref{RicciLinfty}, imply
$$\sum_{i=0}^1||\nabla^i\psi||_{L^\infty(S_{u,\ub})}\leq C(\mathcal O_0, \mathcal R_0, \mathcal G_0).$$
Putting all these together, we get the conclusion of the Proposition.
\end{proof}

In order to close the energy estimates, we will need $L^2$ estimates of up to $3$ derivatives of the Ricci coefficients. The proof of this required $L^2$ estimate is analogous to that in the previous Proposition, except for $\nabla^3\chih, \nabla^3 \trch$ since for that we will need estimates for $\nabla^4\eta$. This can in principle be achievable by elliptic estimates. However, we will take another route. We will use instead the $\nabla_4$ equation for $\nabla^3\trch$ and retrieve the estimates for $\nabla^3\chih$ through elliptic estimates. At this level of derivatives, the $\nabla_4$ equation looks linear in the highest order term. Therefore, we can still achieve estimates that depends only on the initial data. Notice that, however, we cannot control $\nabla^3(\eta,\omegab)$ by the initial data and $\mathcal R(S)$ alone, but need to use $\mathcal R$ as well.
\begin{proposition}\label{Ricci3L2}
Assume $$\mathcal R<\infty.$$
Then there exists $$\epsilon_0=\epsilon_0(\mathcal O_0,\mathcal R_0,\mathcal R, \mathcal G_0)$$ such that for $\epsilon\leq \epsilon_0$, we have
$$\sup_{u,\ub}||\nabla^3(\chibh,\trchb,\etab,\omega)||_{L^2(S_{u,\ub})}\leq 3\mathcal O_0, $$
$$\sup_{u,\ub}||\nabla^3(\chih,\trch)||_{L^2(S_{u,\ub})}\leq C(\mathcal O_0,\mathcal R_0, \mathcal G_0), $$
$$\sup_{u,\ub}||\nabla^3(\eta,\omegab)||_{L^2(S_{u,\ub})}\leq C(\mathcal O_0,\mathcal R_0,\mathcal R, \mathcal G_0).$$
\end{proposition}
\begin{proof}
We introduce the bootstrap assumptions:
$$\sup_{u,\ub}||\nabla^3(\chibh,\trchb,\etab,\omega)||_{L^2(S_{u,\ub})}\leq 4\mathcal O_0,\quad \sup_{u,\ub}||\nabla^3\eta||_{L^2(S_{u,\ub})}\leq \Delta, $$
where $\Delta$ is large and will be chosen later. We take $\epsilon\Delta\ll 1.$
We first estimate $\nabla^3\chih$ and $\nabla^3 \trch$. Unlike in the previous Propositions, we no longer use the $\nabla_3$ equations because of the loss in derivative. Instead we will use the $\nabla_4$ transport equation for $\nabla^3 \trch$ together with the Codazzi equation. Commute the equation for $\nabla_4 \trch$ with $\nabla^3$, we get
$$\nabla_4(\nabla^3 \trch)=\sum_{i_1+i_2+i_3=3}\nabla^{i_1}\psi^{i_2}\nabla^{i_3}((\trch)^2+\chih^2) +\sum_{i_1+i_2+i_3+i_4=3}\nabla^{i_1}\psi^{i_2}\nabla^{i_3}\chi\nabla^{i_4} \trch$$
Notice that in this equation, the only $\nabla^3$ derivatives occur in the terms $\nabla^3 \trch$ and $\nabla^3 \chih$. Hence in order to estimate these terms, we do not need to use the bounds in bootstrap assumption. (We do, however, still need to use the bootstrap assumption in order to apply the previous Propositions.) Now, the right hand side of the equation consists of three kinds of terms. In the first situation, we have $\nabla^3 \trch$ or $\nabla^3\chih$. Then all the other terms do not have derivatives and can be estimated in $L^\infty(S_{u,\ub})$ by $C(\mathcal O_0,\mathcal R_0, \mathcal G_0)$ by Proposition \ref{Ricci}:
$$||\psi\nabla^3\chi||_{L^2(S_{u,\ub'})}\leq ||\psi||_{L^\infty(S_{u,\ub'})})||\nabla^3(\trch,\chih)||_{L^2(S_{u,\ub'})}\leq C(\mathcal O_0,\mathcal R_0, \mathcal G_0)||\nabla^3 (\chih,\trch)||_{L^2(S_{u,\ub'})}.$$
In the second situation, the term has one factor of $\nabla^2\psi$. Then the other factors have at most one derivative. Then, we can estimate using Proposition \ref{Ricci}:
$$||\nabla\psi\nabla^2\psi+\psi\psi\nabla^2\psi||_{L^2(S_{u,\ub'})}\leq (||\nabla\psi||_{L^\infty(S_{u,\ub'})}+||\psi||^2_{L^\infty(S_{u,\ub'})})||\nabla^2\psi||_{L^2(S_{u,\ub'})}\leq C(\mathcal O_0,\mathcal R_0, \mathcal G_0).$$
For the third kind of terms, all factors can be controlled in $L^\infty$:
$$||\psi(\nabla\psi)^2+\psi^3\nabla\psi+\psi^5||_{L^2(S_{u,\ub'})}\leq C(\mathcal O_0,\mathcal R_0, \mathcal G_0).$$
Using Proposition \ref{transport}, we can estimate $\nabla^3 \trch$ by 
$$||\nabla^3 \trch||_{L^2(S_{u,\ub})}\leq C(\mathcal O_0,\mathcal R_0, \mathcal G_0)+C(\mathcal O_0,\mathcal R_0, \mathcal G_0)\int_0^{\ub} ||\nabla^3 (\chih,\trch)||_{L^2(S_{u,\ub'})}d\ub'.$$
By Gronwall's inequality, we have
$$||\nabla^3 \trch||_{L^2(S_{u,\ub})}\leq C(\mathcal O_0,\mathcal R_0, \mathcal G_0)+C(\mathcal O_0,\mathcal R_0, \mathcal G_0)\int_0^{\ub} ||\nabla^3 \chih||_{L^2(S_{u,\ub'})}d\ub'.$$
Using Proposition \ref{ellipticchih} for the Codazzi equation
$$\div\chih=\frac 12 \nabla \trch - \frac 12 (\eta-\etab)\cdot (\chih -\frac 1 2 \trch) -\beta,$$
we have
$$||\nabla^3\chih||_{L^2(S_{u,\ub})}\leq C||\nabla^3 \trch||_{L^2(S_{u,\ub})}+C||\nabla^2\beta||_{L^2(S_{u,\ub})}+C(\mathcal O_0,\mathcal R_0, \mathcal G_0).$$
Notice that we need the estimates for $K$ to apply Proposition \ref{ellipticchih}. This is provided by Proposition \ref{Kest}.
Plugging in the estimate for $||\nabla^3 \trch||_{L^2(S_{u,\ub})}$ as well as the estimate for $\beta$ in Proposition \ref{RS}, we have
$$||\nabla^3\chih||_{L^2(S_{u,\ub})}\leq C(\mathcal O_0,\mathcal R_0, \mathcal G_0)(1+\int_0^{\ub} ||\nabla^3 \chih||_{L^2(S_{u,\ub'})}d\ub').$$
Thus, by Gronwall's inequality, we have
$$||\nabla^3\chih||_{L^2(S_{u,\ub})}\leq C(\mathcal O_0,\mathcal R_0, \mathcal G_0).$$
We now estimate the $L^2$ norms of $\nabla^3\eta$ and $\nabla^3\omegab$. These are analogous to the proof of the previous Proposition, except that now the estimates depend also on $\mathcal R$. This can be improved using elliptic estimates, but since such improvements will not be used later, we will be content with the dependence on $\mathcal R$.
Using the $L^\infty$ bounds for $\psi$ and $\nabla\psi$ and the $L^4$ bounds for $\nabla^2\psi$, we have
$$||\nabla^3\eta||_{L^2(S_{u,\ub})}\leq C(\mathcal O_0,\mathcal R_0,\mathcal R, \mathcal G_0)+C(\mathcal O_0,\mathcal R(S), \mathcal G_0)\int_0^{\ub} ||\nabla^3\eta||_{L^2(S_{u,\ub'})}d\ub'.$$
By Gronwall's inequality,
$$||\nabla^3\eta||_{L^2(S_{u,\ub})}\leq C(\mathcal O_0,\mathcal R_0,\mathcal R, \mathcal G_0).$$
Notice that the right hand side does not depend on $\Delta$. We can therefore choose $\Delta$ large so that this estimate improves the bootstrap assumption.
We then move to the estimate for the $L^2$ norm of $\nabla^3\omegab$. 
$$||\nabla^3\omegab||_{L^2(S_{u,\ub})}\leq C(\mathcal O_0,\mathcal R_0,\mathcal R, \mathcal G_0)+C(\mathcal O_0,\mathcal R_0, \mathcal G_0)\int_0^{\ub} ||\nabla^3\omegab||_{L^2(S_{u,\ub'})}d\ub'.$$
As for the estimates for $\nabla\eta$, we use Gronwall's inequality to get
$$||\nabla^3\omegab||_{L^2(S_{u,\ub})}\leq C(\mathcal O_0,\mathcal R_0,\mathcal R, \mathcal G_0).$$
We now move to close the rest of the bootstrap assumptions:
$$||\nabla^3(\chibh,\trchb,\etab,\omega)||_{L^2(S_{u,\ub})}\leq 2\mathcal O_0+C\epsilon^{\frac{1}{2}}\mathcal R+\epsilon C(\mathcal O_0,\mathcal R_0, \mathcal G_0)\sup_{\ub'\in [0,\ub]}||\nabla^3(\chibh,\trchb,\etab,\omega)||_{L^2(S_{u,\ub'})}.$$
Hence, by choosing $\epsilon$ sufficiently small, we have
$$||\nabla^3(\chibh,\trchb,\etab,\omega)||_{L^2(S)}\leq 3\mathcal O_0,$$
which improves the bootstrap assumption and gives the desired estimates.
\end{proof}
We now gather all the estimates for the Ricci coefficients that we have obtained:
\begin{proposition}\label{Ricci2}
Assume
$$\mathcal R<\infty.$$
Then there exists $\epsilon_0=\epsilon_0(\mathcal O_0, \mathcal R(S),\displaystyle\sup_{u,\ub}||\nabla^3\eta||_{L^2(S_{u,\ub})},\mathcal R, \mathcal G_0)$ such that for $\epsilon\leq \epsilon_0$, we have
$$\sup_{u,\ub}\sup_{\{\psi\in\chih,\trch,\chibh,\trchb,\etab,\omega\}}(\sum_{i=0}^1||\nabla^i\psi||_{L^\infty(S_{u,\ub})}+\sum_{i=0}^2||\nabla^i \psi||_{L^4(S_{u,\ub})}+\sum_{i=0}^3||\nabla^i\psi||_{L^2(S_{u,\ub})})\leq C(\mathcal O_0,\mathcal R_0, \mathcal G_0), $$
$$\sup_{u,\ub}(||(\eta,\omegab)||_{L^2(S_{u,\ub})}+\sum_{i=0}^1||\nabla^i(\eta,\omegab)||_{L^2(S_{u,\ub})}+\sum_{i=0}^2||\nabla^i(\eta,\omegab)||_{L^2(S_{u,\ub})})
\leq C(\mathcal O_0,\mathcal R_0,\mathcal G_0), $$
$$\sup_{u,\ub}(||\nabla(\eta,\omegab)||_{L^2(S_{u,\ub})}+||\nabla^2(\eta,\omegab)||_{L^2(S_{u,\ub})}+||\nabla^3(\eta,\omegab)||_{L^2(S_{u,\ub})})\leq C(\mathcal O_0,\mathcal R_0,\mathcal R, \mathcal G_0).$$
\end{proposition}
\begin{proof}
This is a direct consequence of Proposition \ref{Ricci}, \ref{Ricci3L2} and the Sobolev Embedding in Propositions \ref{L4} and \ref{Linfty}.
\end{proof}
\begin{remark}
In particular, we have recovered the bootstrap assumption (\ref{BA}).
\end{remark}

We now begin step two of the proof, i.e., the energy type estimates for the curvature components. In order to derive these estimates, we need some integration by parts identities. We first show the following integration by parts formula for the $u,\ub$ variables.
\begin{proposition}\label{intbyparts34}
Suppose $\phi_1$ and $\phi_2$ are $r$ tensorfields, then
$$\int_{D_{u,\ub}} \phi_1 \nabla_4\phi_2+\int_{D_{u,\ub}}\phi_2\nabla_4\phi_1= \int_{\Hb_{\ub}(0,u)} \phi_1\phi_2-\int_{\Hb_0(0,u)} \phi_1\phi_2+\int_{D_{u,\ub}}(2\omega-\frac 12\trch)\phi_1\phi_2,$$
$$\int_{D_{u,\ub}} \phi_1 \nabla_3\phi_2+\int_{D_{u,\ub}}\phi_2\nabla_3\phi_1= \int_{H_{u}(0,\ub)} \phi_1\phi_2-\int_{H_0(0,\ub)} \phi_1\phi_2+\int_{D_{u,\ub}}(2\omegab-\frac 12\trchb)\phi_1\phi_2.$$
\end{proposition}
\begin{proof}
To prove the second estimate, we use the coordinates $(u,\ub,\theta^1,\theta^2)$. The first estimate follows identically by consider instead the $(u,\ub,\thb^1,\thb^2)$ coordinates.
For tensorfields $\phi_1$ and $\phi_2$, $\phi_1\phi_2$ is understood to be the pairing using the metric $\gamma$. Since $\nabla_3\gamma=0$, we have
$$\phi_1 \nabla_3\phi_2=\nabla_3 (\phi_1\phi_2)-\phi_2\nabla_3\phi_1.$$
In the $(u,\ub,\th^1,\th^2)$ coordinates, the volume form is
$$\Omega^2 \sqrt{\det\gamma}.$$
Now, $f=\phi_1\phi_2$ is a scalar function.
\begin{equation*}
\begin{split}
&\int_{D_{u,\ub}} \nabla_3 f\\
=&\sum_U\int_0^u\int_0^{\ub}\int_{-\infty}^{\infty} \frac{\partial f}{\partial u}\Omega\sqrt{\det\gamma} d\th^1d\th^2d\ub' du'\\
=&\sum_U\int_0^{\ub}\int_{-\infty}^{\infty} f\Omega\sqrt{\det\gamma}(u) d\th^1d\th^2d\ub' du'-\sum_U\int_0^{\ub}\int_{-\infty}^{\infty} f\Omega\sqrt{\det\gamma}(u=0) d\th^1d\th^2d\ub' du'\\
&+\sum_U\int_0^u\int_0^{\ub}\int_{-\infty}^{\infty} f(2\omega-\frac 12\trch)\Omega^2\sqrt{\det\gamma} d\th^1d\th^2d\ub' du'\\
=&\int_{H_u(0,\ub)}f-\int_{H_0(0,\ub)}f+\int_{D_{u,\ub}} f(2\omega-\frac 12\trchb),
\end{split}
\end{equation*}
where we have used $\frac{\partial}{\partial u}\log(\det\gamma)=\Omega\trch$ and $\nabla_3\log\Omega=-2\omega.$
\end{proof}
We also have the following integration by parts formula on the 2-sphere. 
\begin{proposition}\label{intbypartssph}
Suppose we have an $r$ tensorfield $^{(1)}\phi$ and an $r-1$ tensorfield $^{(2)}\phi$.
$$\int_{D_{u,\ub}}{ }^{(1)}\phi^{A_1A_2...A_r}\nabla_{A_r}{ }^{(2)}\phi_{A_1...A_{r-1}}+\int_{D_{u,\ub}}\nabla^{A_r}{ }^{(1)}\phi_{A_1A_2...A_r}{ }^{(2)}\phi^{A_1...A_{r-1}}= -\int_{D_{u,\ub}}(\eta+\etab){ }^{(1)}\phi{ }^{(2)}\phi.$$
\end{proposition}
\begin{proof}
First, since $\nabla\gamma=0$, the following is simply integration by parts on the 2-sphere:
\begin{equation*}
\begin{split}
\int_{S_{u,\ub}}\gamma^{A_1B_1}...\gamma^{A_r B_r}{ }^{(1)}\phi_{A_1A_2...A_r}\nabla_{B_r}{ }^{(2)}\phi_{B_1...B_{r-1}}
=&-\int_{S_{u,\ub}}\gamma^{A_1B_1}...\gamma^{A_r B_r}\nabla_{B_r}{ }^{(1)}\phi_{A_1A_2...A_r}{ }^{(2)}\phi_{B_1...B_{r-1}}.
\end{split}
\end{equation*}
The difference when we integrate over $D_{u,\ub}$ instead of $S_{u,\ub}$ is in the volume form:
\begin{equation*}
\begin{split}
&\int_{D_{u,\ub}}\gamma^{A_1B_1}...\gamma^{A_r B_r}{ }^{(1)}\phi_{A_1A_2...A_r}\nabla_{B_r}{ }^{(2)}\phi_{B_1...B_{r-1}}\\
=&\int_0^u\int_0^{\ub}\int_{S_{u',\ub'}}\gamma^{A_1B_1}...\gamma^{A_r B_r}{ }^{(1)}\phi_{A_1A_2...A_r}\nabla_{B_r}{ }^{(2)}\phi_{B_1...B_{r-1}} \Omega^2 d\ub' du'\\
=&-\int_{D_{u,\ub}}\gamma^{A_1B_1}...\gamma^{A_r B_r}\nabla_{B_r}{ }^{(1)}\phi_{A_1A_2...A_r}{ }^{(2)}\phi_{B_1...B_{r-1}}-\int_{D_{u,\ub}}{ }^{(1)}\phi_{A_1A_2...A_r}{ }^{(2)}\phi^{A_1...A_{r-1}}\nabla^{A_r}(\log\Omega^2)\\
=&-\int_{D_{u,\ub}}\gamma^{A_1B_1}...\gamma^{A_r B_r}\nabla_{B_r}{ }^{(1)}\phi_{A_1A_2...A_r}{ }^{(2)}\phi_{B_1...B_{r-1}}-\int_{D_{u,\ub}}{ }^{(1)}\phi_{A_1A_2...A_r}{ }^{(2)}\phi^{A_1...A_{r-1}}(\eta^{A_r}+\etab^{A_r}).
\end{split}
\end{equation*}
In the above, we have used 
$$\nabla^{A_r}(\log\Omega^2)=(\eta^{A_r}+\etab^{A_r}).$$
\end{proof}
Finally, we prove the energy estimates. We will derive the energy estimates by directly using the Bianchi equations and the integration by parts formulae in Proposition \ref{intbyparts34} and \ref{intbypartssph}.
\begin{proposition}\label{ee}
The following $L^2$ estimates for the curvature hold:
\begin{equation*}
\begin{split}
&\sum_{\Psi\in\{\alpha,\beta,\rho,\sigma\}}\int_{H_u} |\nabla^i\Psi|^2_\gamma+\sum_{\Psi\in\{\beta,\rho,\sigma,\betab\}}\int_{\Hb_{\ub}} |\nabla^i\Psi|^2_\gamma  \\
\leq &\sum_{\Psi\in\{\alpha,\beta,\rho,\sigma\}}\int_{H_{u'}} |\nabla^i\Psi|^2_\gamma+\sum_{\Psi\in\{\rho,\sigma,\betab\}}\int_{\Hb_{\ub'}} |\nabla^i\Psi|^2_\gamma \\
&+\sum_{\substack{\Psi_1\in\{\alpha,\beta,\rho,\sigma,\betab\}\\ \Psi_2\in\{\alpha,\beta,\rho,\sigma,\betab,\alphab\}}}\int_{D_{u,\ub}}\nabla^i\Psi_1\sum_{i_1+i_2+i_3+i_4=i}\nabla^{i_1}\psi^{i_2}\nabla^{i_3}\psi\nabla^{i_4}\Psi_{2}\\
&+\sum_{\substack{\Psi_1\in\{\alpha,\beta,\rho,\sigma,\betab\}\\\Psi_2\in\{\alpha,\beta,\rho,\sigma,\betab,\alphab\}}}\int_{D_{u,\ub}}\nabla^i\Psi_1\sum_{i_1+i_2+i_3+i_4=i-1}\nabla^{i_1}\psi^{i_2}\nabla^{i_3}K\nabla^{i_4}\Psi_{2}.
\end{split}
\end{equation*}

\end{proposition}
\begin{proof}
\begin{equation*}
\begin{split}
&\nabla_3\alpha+\frac 12 \trchb \alpha=\nabla\hot \beta+ 4\omegab\alpha-3(\chih\rho+^*\chih\sigma)+
(\zeta+4\eta)\hot\beta,\\
&\nabla_4\beta+2\trch\beta = \div\alpha - 2\omega\beta +  \eta \alpha,\\
\end{split}
\end{equation*}
Schematically, we can write
\begin{equation*}
\begin{split}
&\nabla_3\alpha=\nabla\hot \beta+\psi\Psi,\\
&\nabla_4\beta = \div\alpha +\psi\Psi.\\
\end{split}
\end{equation*}
Now, applying integration by parts by Proposition \ref{intbypartssph} , we have
\begin{equation*}
\begin{split}
\int <\alpha,\nabla_3\alpha>_\gamma =&\int <\alpha,\nabla\hot\beta>_\gamma+<\alpha,\psi\Psi>_\gamma \\
=&\int -<\div\alpha,\beta>_\gamma+<\alpha,\psi\Psi>_\gamma \\
=&\int -<\nabla_4\beta,\beta>_\gamma+<\alpha,\psi\Psi>_\gamma +<\beta,\psi\Psi>_\gamma\\
\end{split}
\end{equation*}
Apply the integration by parts given by Proposition \ref{intbyparts34} to get that for $u\geq u'$, $\ub\geq \ub'$,
\begin{equation*}
\begin{split}
\int_{H_u} |\alpha|^2_\gamma+\int_{\Hb_{\ub}} |\beta|^2_\gamma \leq \int_{H_{u'}} |\alpha|^2_\gamma+\int_{\Hb_{\ub'}} |\beta|^2_\gamma+\int_{D_{u,\ub}}<(\alpha,\beta),\psi\Psi>_\gamma\\
\end{split}
\end{equation*}
We use the commutation formula, and note that the special structure is preserved in the highest order:
\begin{equation*}
\begin{split}
&\nabla_3\nabla^i\alpha- \nabla\hot \nabla^i\beta \\
\sim&\sum_{i_1+i_2+i_3+i_4=i}\nabla^{i_1}\psi^{i_2}\nabla^{i_3}\chibh\nabla^{i_4}\alpha \\
&+\sum_{i_1+i_2+i_3+i_4+i_5=i-1}\nabla^{i_1}\psi^{i_2}\nabla^{i_3}K^{i_4}\nabla^{i_5}\beta+\sum_{i_1+i_2+i_3=i}\nabla^{i_1}\psi^{i_2}\nabla^{i_3}(\psi\Psi),\\
&\nabla_4\nabla^i\beta- \div\nabla^i\alpha\\
\sim&\sum_{i_1+i_2+i_3+i_4=i}\nabla^{i_1}\psi^{i_2}\nabla^{i_3}\chih\nabla^{i_4}\beta \\
&+\sum_{i_1+i_2+i_3+i_4+i_5=i-1}\nabla^{i_1}\psi^{i_2}\nabla^{i_3}K^{i_4}\nabla^{i_5}\alpha+\sum_{i_1+i_2+i_3=i}\nabla^{i_1}\psi^{i_2}\nabla^{i_3}(\psi\Psi),\\
\end{split}
\end{equation*}
where $$(\nabla\hot \nabla^i\beta)_{BA_1...A_iC}=(\nabla^{i+1}\beta)_{BA_1...A_iC}+(\nabla^{i+1}\beta)_{CA_1...A_iB}-\gamma_{BC}(\nabla^{i+1}\beta)^D{ }_{A_1...A_iD},$$
$$(\div\nabla^i\alpha)_{A_1...A_iB}=(\nabla^{i+1}\alpha)^C{ }_{A_1...A_iBC}.$$
Perform integration by parts using Proposition \ref{intbypartssph} and \ref{intbyparts34} as before,
\begin{equation*}
\begin{split}
&\int_{H_u} |\nabla^i\alpha|^2_\gamma+\int_{\Hb_{\ub}} |\nabla^i\beta|^2_\gamma  \\
\leq &\int_{H_{u'}} |\nabla^i\alpha|^2_\gamma+\int_{\Hb_{\ub'}} |\nabla^i\beta|^2_\gamma+\int_{D_{u,\ub}}\nabla^i(\alpha,\beta)\sum_{i_1+i_2+i_3+i_4=i}\nabla^{i_1}\psi^{i_2}\nabla^{i_3}\psi\nabla^{i_4}\Psi\\
&+\int_{D_{u,\ub}}\nabla^i(\alpha,\beta)\sum_{i_1+i_2+i_3+i_4=i}\nabla^{i_1}(\eta+\etab)^{i_2}\nabla^{i_3}K\nabla^{i_4}\Psi.
\end{split}
\end{equation*}
We then consider the following sets of Bianchi equations:
\begin{equation*}
\begin{split}
&\nabla_3\beta+\trchb\beta=\nabla\rho + 2\omegab \beta +^*\nabla\sigma +2\chih\cdot\betab+3(\eta\rho+^*\eta\sigma),\\
&\nabla_4\sigma+\frac 32\trch\sigma=-\div^*\beta+\frac 12\chibh\cdot ^*\alpha-\zeta\cdot^*\beta-2\etab\cdot
^*\beta,\\
&\nabla_4\rho+\frac 32\trch\rho=\div\beta-\frac 12\chibh\cdot\alpha+\zeta\cdot\beta+2\etab\cdot\beta,\\
\end{split}
\end{equation*}
Applying Proposition \ref{intbypartssph} and \ref{intbyparts34} yields the following energy estimates on the $0$-th derivative of the curvature.
\begin{equation*}
\begin{split}
\int <\beta,\nabla_3\beta>_\gamma =&\int <\beta,\nabla\rho+^*\nabla\sigma>_\gamma+<\beta,\psi\Psi>_\gamma \\
=&\int -<\div\beta,\rho>_\gamma+<\div ^*\beta,\sigma>_\gamma +<\beta,\psi\Psi>_\gamma\\
=&\int -<\nabla_4\rho,\rho>_\gamma-<\nabla_4\sigma,\sigma>_\gamma +<(\beta,\rho,\sigma),\psi\Psi>_\gamma\\
\end{split}
\end{equation*}
As before, we commute the Bianchi equations with $\nabla^i$ and derive the following:
\begin{equation*}
\begin{split}
&\int_{H_u} |\nabla^i\beta|^2_\gamma+\int_{\Hb_{\ub}} |\nabla^i(\rho,\sigma)|^2_\gamma  \\
\leq &\int_{H_{u'}} |\nabla^i\beta|^2_\gamma+\int_{\Hb_{\ub'}} |\nabla^i(\rho,\sigma)|^2_\gamma+\int_{D_{u,\ub}}\nabla^i(\beta,\rho,\sigma)\sum_{i_1+i_2+i_3+i_4=i}\nabla^{i_1}\psi^{i_2}\nabla^{i_3}\psi\nabla^{i_4}\Psi\\
&+\int_{D_{u,\ub}}\nabla^i(\beta,\rho,\sigma)\sum_{i_1+i_2+i_3+i_4=i}\nabla^{i_1}\psi^{i_2}\nabla^{i_3}K\nabla^{i_4}\Psi.
\end{split}
\end{equation*}
We then consider the following set of Bianchi equations. They are ``symmetric'' to the previous set by changing 3 and 4 appropriately.
\begin{equation*}
\begin{split}
&\nabla_3\sigma+\frac 32\trchb\sigma=-\div ^*\betab+\frac 12\chih\cdot ^*\alphab-\zeta\cdot ^*\betab-2\eta\cdot 
^*\betab,\\
&\nabla_3\rho+\frac 32\trchb\rho=-\div\betab- \frac 12\chih\cdot\alphab+\zeta\cdot\betab-2\eta\cdot\betab,\\
&\nabla_4\betab+\trch\betab=-\nabla\rho +^*\nabla\sigma+ 2\omega\betab +2\chibh\cdot\beta-3(\etab\rho-^*\etab\sigma),\\
\end{split}
\end{equation*}
From these we can derive the following estimates:
\begin{equation*}
\begin{split}
&\int_{H_u} |\nabla^i(\rho,\sigma)|^2_\gamma+\int_{\Hb_{\ub}} |\nabla^i\betab|^2_\gamma  \\
\leq &\int_{H_{u'}} |\nabla^i(\rho,\sigma)|^2_\gamma+\int_{\Hb_{\ub'}} |\nabla^i\betab|^2_\gamma+\int_{D_{u,\ub}}\nabla^i(\rho,\sigma,\betab)\sum_{i_1+i_2+i_3+i_4=i}\nabla^{i_1}\psi^{i_2}\nabla^{i_3}\psi\nabla^{i_4}\Psi\\
&+\int_{D_{u,\ub}}\nabla^i(\rho,\sigma,\betab)\sum_{i_1+i_2+i_3+i_4=i}\nabla^{i_1}\psi^{i_2}\nabla^{i_3}K\nabla^{i_4}\Psi.
\end{split}
\end{equation*}
\end{proof}
We also have the following estimates for $\betab$ in $L^2(H_u)$ and for $\alphab$ in $L^2(\Hb_{\ub})$. The key observation is that for terms that contain $\nabla^i\alphab\nabla^j\alphab$, $\nabla^3\eta$ and $\nabla^3\omega$ do not appear. The proof is exactly as in the proof of the last Proposition. We will nevertheless go through the details again to show the relevant terms.
\begin{proposition}\label{ee2}
\begin{equation*}
\begin{split}
&\int_{H_u} |\nabla^i\betab|^2_\gamma+\int_{\Hb_{\ub}} |\nabla^i\alphab|^2_\gamma  \\
\leq &\int_{H_{u'}} |\nabla^i\betab|^2_\gamma+\int_{\Hb_{\ub'}} |\nabla^i\alphab|^2_\gamma +\int_{D_{u,\ub}} \nabla^i\alphab\sum_{i_1+i_2+i_3+i_4=i}\nabla^{i_1}\psi^{i_2}(\nabla^{i_3}\omega+\nabla^{i_3}\chi)\nabla^{i_4}\alphab\\
&+\int_{D_{u,\ub}} \nabla^i\alphab\sum_{i_1+i_2+i_3+i_4=i-1}\nabla^{i_1}\psi^{i_2}\nabla^{i_3}K\nabla^{i_4}\alphab\\
&+\sum_{\substack{\Psi_{H}\in\{\alpha,\beta,\rho,\sigma,\betab\}\\ \Psi\in\{\alpha,\beta,\rho,\sigma,\betab,\alphab\}}}\int_{D_{u,\ub}}\nabla^i\Psi_{H}\sum_{i_1+i_2+i_3+i_4=i}\nabla^{i_1}\psi^{i_2}\nabla^{i_3}\psi\nabla^{i_4}\Psi\\
&+\sum_{\substack{\Psi_{H}\in\{\alpha,\beta,\rho,\sigma,\betab\}\\ \Psi\in\{\alpha,\beta,\rho,\sigma,\betab,\alphab\}}}\int_{D_{u,\ub}}\nabla^i\Psi\sum_{i_1+i_2+i_3+i_4=i}\nabla^{i_1}\psi^{i_2}\nabla^{i_3}\psi\nabla^{i_4}\Psi_H\\
&+\sum_{\substack{\Psi_{H}\in\{\alpha,\beta,\rho,\sigma,\betab\}\\ \Psi\in\{\alpha,\beta,\rho,\sigma,\betab,\alphab\}}}\int_{D_{u,\ub}}\nabla^i\Psi_{H}\sum_{i_1+i_2+i_3+i_4=i-1}\nabla^{i_1}\psi^{i_2}\nabla^{i_3}K\nabla^{i_4}\Psi\\
&+\sum_{\substack{\Psi_{H}\in\{\alpha,\beta,\rho,\sigma,\betab\}\\ \Psi\in\{\alpha,\beta,\rho,\sigma,\betab,\alphab\}}}\int_{D_{u,\ub}}\nabla^i\Psi\sum_{i_1+i_2+i_3+i_4=i-1}\nabla^{i_1}\psi^{i_2}\nabla^{i_3}K\nabla^{i_4}\Psi_H\\
\end{split}
\end{equation*}
\end{proposition}
\begin{proof}
We look at the following set of Bianchi equations:
\begin{equation*}
\begin{split}
&\nabla_3\betab+2\trchb\betab=-\div\alphab-2\omegab\betab+\etab \cdot\alphab,\\
&\nabla_4\alphab+\frac 12 \trch\alphab=-\nabla\hot \betab+ 4\omega\alphab-3(\chibh\rho-^*\chibh\sigma)+
(\zeta-4\etab)\hot \betab
\end{split}
\end{equation*}
In schematic notation, noting the $\alphab$ terms, we have
\begin{equation*}
\begin{split}
&\nabla_3\betab+\div\alphab\sim \psi\Psi,\\
&\nabla_4\alphab+\nabla\hot \betab\sim \psi\Psi_H+\trch\alphab+ \omega\alphab,
\end{split}
\end{equation*}
where as in the statement of the Proposition, we use $\Psi_H$ to denote null curvature components that are not $\alphab$.
Commuting the equations with up to $3$ angular derivatives, we get
\begin{equation*}
\begin{split}
&\nabla_3\nabla^i\betab- \div \nabla^i\alphab \\
\sim&\sum_{i_1+i_2+i_3+i_4=i}\nabla^{i_1}\psi^{i_2}\nabla^{i_3}\chibh\nabla^{i_4}\betab\\
&+\sum_{i_1+i_2+i_3+i_4+i_5=i-1}\nabla^{i_1}\psi^{i_2}\nabla^{i_3}K^{i_4}\nabla^{i_5}\alphab+\sum_{i_1+i_2+i_3=i}\nabla^{i_1}\psi^{i_2}\nabla^{i_3}(\psi\Psi)=:F_1,\\
&\nabla_4\nabla^i\alphab- \nabla\hot\nabla^i\betab\\
\sim&\sum_{i_1+i_2+i_3+i_4=i}\nabla^{i_1}(\eta+\etab)^{i_2}\nabla^{i_3}\chih\nabla^{i_4}\alphab\\
&+\sum_{i_1+i_2+i_3+i_4+i_5=i-1}\nabla^{i_1}\psi^{i_2}\nabla^{i_3}K^{i_4}\nabla^{i_5}\betab+\sum_{i_1+i_2+i_3=i}\nabla^{i_1}\psi^{i_2}\nabla^{i_3}(\psi\Psi_H)\\
&+\sum_{i_1+i_2+i_3=i}\nabla^{i_1}\psi^{i_2}\nabla^{i_3}(\trch\alphab+\omega\alphab)=:F_2,
\end{split}
\end{equation*}
where $\Psi_H\in\{\alpha,\beta,\rho,\sigma,\betab\}$.
Using these two identities, we get
\begin{equation*}
\begin{split}
&\int_{H_u}|\nabla^i\betab|^2_\gamma-\int_{H_0}|\nabla^i\betab|^2_\gamma+ \int_{\Hb_{\ub}}|\nabla^i\alphab|^2_{\gamma} -\int_{\Hb_{0}}|\nabla^i\alphab|^2_{\gamma}\\
\leq &\int_{D_{u,\ub}}(\nabla^i\betab\nabla_3\nabla^i\betab- \nabla^i\alphab \nabla_4\nabla^i\alphab)+\int_{D_{u,\ub}}(|\omegab|+|\trchb|)|\nabla^i\betab|_{\gamma}^2+\int_{D_{u,\ub}}(|\omega|+|\trch|)|\nabla^i\alphab|_{\gamma}^2\\
\leq &\int_{D_{u,\ub}}\nabla^i\betab\nabla_3\nabla^i\betab- \nabla^i\betab \div \nabla^i\alphab+\int_{D_{u,\ub}}\nabla^i\alphab\nabla_4\nabla^i\alphab- \nabla^i\alphab \nabla\hot \nabla^i\betab\\
&+\int_{D_{u,\ub}}(|\omegab|+|\trchb|)|\nabla^i\betab|_{\gamma}^2+\int_{D_{u,\ub}}(|\omega|+|\trch|)|\nabla^i\alphab|_{\gamma}^2+\int_{D_{u,\ub}}|\zeta|_{\gamma}|\nabla^i\alphab|_{\gamma}|\nabla^2\betab|_{\gamma}\\
\leq &\int_{D_{u,\ub}}F_1\nabla^i\betab+\int_{D_{u,\ub}}F_2\nabla^i\alphab\\
&+\int_{D_{u,\ub}}(|\omegab|+|\trchb|)|\nabla^i\betab|_{\gamma}^2+\int_{D_{u,\ub}}(|\omega|+|\trch|)|\nabla^i\alphab|_{\gamma}^2+\int_{D_{u,\ub}}|\zeta|_{\gamma}|\nabla^i\alphab|_{\gamma}|\nabla^2\betab|_{\gamma}.\\
\end{split}
\end{equation*}
The conclusion now follows from the structure of $F_1$ and $F_2$.
\end{proof}

Using the above Propositions, we can show the boundedness of the curvature component:
\begin{proposition}\label{finalcurvbound}
$$\mathcal R\leq C(\mathcal O_0,\mathcal R_0, \mathcal G_0).$$
\end{proposition}
\begin{proof}
We have to control the terms integrated in $D_{u,\ub}$ in Propositions \ref{ee} and \ref{ee2} for $i\leq 3$. We first control the terms from Proposition \ref{ee}. We need to control
\begin{equation}\label{eegood}
\begin{split}
&\sum_{\substack{\Psi_1\in\{\alpha,\beta,\rho,\sigma,\betab\}\\ \Psi_2\in\{\alpha,\beta,\rho,\sigma,\betab,\alphab\}}}\int_{D_{u,\ub}}\nabla^i\Psi_1\sum_{i_1+i_2+i_3+i_4=i}\nabla^{i_1}\psi^{i_2}\nabla^{i_3}\psi\nabla^{i_4}\Psi_{2}\\
&+\sum_{\substack{\Psi_1\in\{\alpha,\beta,\rho,\sigma,\betab\}\\\Psi_2\in\{\alpha,\beta,\rho,\sigma,\betab,\alphab\}}}\int_{D_{u,\ub}}\nabla^i\Psi_1\sum_{i_1+i_2+i_3+i_4=i-1}\nabla^{i_1}\psi^{i_2}\nabla^{i_3}K\nabla^{i_4}\Psi_{2}.
\end{split}
\end{equation}
We begin with the first term. Since $\Psi_1\in\{\alpha,\beta,\rho,\sigma,\betab\}$, we can control $\nabla^i\Psi_1$ in $L^2(H_u)$, thus
$$\sum_{i=0}^3||\nabla^i\Psi_1||_{L^2(D_{u,\ub})}\leq C\epsilon^{\frac{1}{2}}\mathcal R.$$
For $\Psi_2\in\{\alpha,\beta,\rho,\sigma,\betab,\alphab\}$, since $\alphab$ can only be controlled in $L^2(\Hb_{\ub})$, we only have
$$\sum_{i=0}^3||\nabla^i\Psi_2||_{L^2(D_{u,\ub})}\leq C\mathcal R.$$
By Cauchy-Schwarz, and using Proposition \ref{Ricci2}, for $\Psi_1\in\{\alpha,\beta,\rho,\sigma,\betab\}$ and $\Psi_2\in\{\alpha,\beta,\rho,\sigma,\betab,\alphab\}$, we have
\begin{equation*}
\begin{split}
&\sum_{i=0}^3||\nabla^i\Psi_1||_{L^2(D_{u,\ub})}\sum_{i_1+i_2+i_3+i_4\leq 3}||\nabla^{i_1}\psi^{i_2}\nabla^{i_3}\psi\nabla^{i_4}\Psi_{2}||_{L^2(D_{u,\ub})}\\
\leq &C\epsilon^{\frac{1}{2}}\mathcal R (1+\sum_{i_1=0}^1||\nabla^{i_1}\psi||_{L^\infty(S_{u,\ub})})^4(1+||\nabla^2\psi||_{L^4(S_{u,\ub})}+||\nabla^3\psi||_{L^2(S_{u,\ub})})\sum_{i_3=0}^3||\nabla^i\Psi_2||_{L^2(D_{u,\ub})}\\
\leq &C(\mathcal O_0,\mathcal R_0,\mathcal R,\mathcal G_0)\epsilon^{\frac{1}{2}}.
\end{split}
\end{equation*}
For the second term we again use Cauchy-Schwarz, and apply Proposition \ref{Ricci2} and \ref{Kest} to get
\begin{equation*}
\begin{split}
&||\nabla^i\Psi_1||_{L^2(D_{u,\ub})}\sum_{i_1+i_2+i_3+i_4\leq 2}||\nabla^{i_1}\psi^{i_2}\nabla^{i_3}K\nabla^{i_4}\Psi_{2}||_{L^2(D_{u,\ub})}\\
\leq &C\epsilon^{\frac{1}{2}}\mathcal R (1+\sum_{i_1=0}^1||\nabla^{i_1}\psi||_{L^\infty(S_{u,\ub})})^2(||K||_{L^\infty(S_{u,\ub})}+||\nabla K||_{L^4(S_{u,\ub})}+||\nabla^2 K||_{L^2(S_{u,\ub})})\sum_{i_2=0}^3||\nabla^i\Psi_2||_{L^2(D_{u,\ub})}\\
\leq &C(\mathcal O_0,\mathcal R_0,\mathcal R,\mathcal G_0)\epsilon^{\frac{1}{2}}.
\end{split}
\end{equation*}
We now move to estimating terms from Proposition \ref{ee2}. We first estimate all the terms that do not involve $\alphab^2$. We claim that all such terms have an extra power of $\epsilon$. The terms
\begin{equation*}
\begin{split}
&\sum_{\substack{\Psi_{H}\in\{\alpha,\beta,\rho,\sigma,\betab\}\\ \Psi\in\{\alpha,\beta,\rho,\sigma,\betab,\alphab\}}}\int_{D_{u,\ub}}\nabla^i\Psi_{H}\sum_{i_1+i_2+i_3+i_4=i}\nabla^{i_1}\psi^{i_2}\nabla^{i_3}\psi\nabla^{i_4}\Psi\\
&+\sum_{\substack{\Psi_{H}\in\{\alpha,\beta,\rho,\sigma,\betab\}\\ \Psi\in\{\alpha,\beta,\rho,\sigma,\betab,\alphab\}}}\int_{D_{u,\ub}}\nabla^i\Psi_{H}\sum_{i_1+i_2+i_3+i_4=i-1}\nabla^{i_1}\psi^{i_2}\nabla^{i_3}K\nabla^{i_4}\Psi
\end{split}
\end{equation*}
are identical to the terms in (\ref{eegood}) and thus can be estimated in the same manner by 
$$C(\mathcal O_0,\mathcal R_0,\mathcal R,\mathcal G_0)\epsilon^{\frac{1}{2}}.$$
The terms
\begin{equation*}
\begin{split}
&\sum_{\substack{\Psi_{H}\in\{\alpha,\beta,\rho,\sigma,\betab\}\\ \Psi\in\{\alpha,\beta,\rho,\sigma,\betab,\alphab\}}}\int_{D_{u,\ub}}\nabla^i\Psi\sum_{i_1+i_2+i_3+i_4=i}\nabla^{i_1}\psi^{i_2}\nabla^{i_3}\psi\nabla^{i_4}\Psi_H\\
&+\sum_{\substack{\Psi_{H}\in\{\alpha,\beta,\rho,\sigma,\betab\}\\ \Psi\in\{\alpha,\beta,\rho,\sigma,\betab,\alphab\}}}\int_{D_{u,\ub}}\nabla^i\Psi\sum_{i_1+i_2+i_3+i_4=i-1}\nabla^{i_1}\psi^{i_2}\nabla^{i_3}K\nabla^{i_4}\Psi_H
\end{split}
\end{equation*}
can also be treated in a similar manner by noting that $\Psi_H$ can be estimated in $L^2(H_u)$. We are thus left with the terms involving $\alphab^2$, i.e.,
$$\int_{D_{u,\ub}} \sum_{i\leq 3,i_1+i_2+i_3+i_4\leq 3}\nabla^i\alphab\nabla^{i_1}\psi^{i_2}\nabla^{i_3}(\omega,\chih,\trch)\nabla^{i_4}\alphab+\sum_{i\leq 3, i_1+i_2+i_3+i_4\leq 2}\nabla^i\alphab\nabla^{i_1}\psi^{i_2}\nabla^{i_3} K\nabla^{i_4}\alphab.$$
By Holder's inequality, this can be estimated by 
$$\int_{0}^{\ub}\sum_{i=0}^3||\nabla^i\alphab||_{L^2(\Hb_{\ub'})}(1+\sum_{l=0}^2||\nabla^l\psi||_{L^4(S_{u,\ub})})^3 (\sum_{j+k\leq 3}||\nabla^j(\omega,\chih,\trch)\nabla^k\alphab||_{L^2(H_{\ub'})}+\sum_{j+k\leq 2}||\nabla^jK\nabla^k\alphab||_{L^2(H_{\ub'})})d\ub'.$$
We first consider the first term. By Proposition \ref{Ricci2}, we have
$$\sum_{l=0}^2||\nabla^l\psi||_{L^4(S_{u,\ub})}\leq C(\mathcal O_0,\mathcal R_0,\mathcal G_0).$$
For $\psi\in\{\omega,\chih,\trch\}$, we have
$$\sum_{j+k\leq 3}||\nabla^j\psi\nabla^k\alphab||_{L^2(H_{\ub'})}\leq C\sum_{k=0}^3||\nabla^k\alphab||_{L^2(\Hb_{\ub'})}\sup_{u,\ub}(||\psi||_{L^\infty(S_{u,\ub})}+\sum_{j=0}^2||\nabla^j\psi||_{L^4(S_{u,\ub})}+||\nabla^3\psi||_{L^2(S_{u,\ub})}).$$
By Proposition \ref{Ricci2}, we can control the term by 
$$C(\mathcal O_0,\mathcal R_0, \mathcal G_0)\int_{0}^{\ub}\sum_{i=0}^3||\nabla^i\alphab||^2_{L^2(\Hb_{\ub})}d\ub'.$$
The second term can be control analogously using Proposition \ref{Kest} by
$$C(\mathcal O_0,\mathcal R_0, \mathcal G_0)\int_{0}^{\ub}\sum_{i=0}^3||\nabla^i\alphab||^2_{L^2(\Hb_{\ub})}d\ub'.$$
So we have
$$\sum_{i=0}^{3} ||\nabla^i\alphab||_{L^2(\Hb_{\ub})}^2\leq C\mathcal R_0+C(\mathcal O_0,\mathcal R)\epsilon^{\frac{1}{2}}+C(\mathcal O_0,\mathcal R_0, \mathcal G_0)\sum_{i=0}^{3}\int_0^{\ub} ||\nabla^i\alphab||_{L^2(\Hb_{\ub'})}^2d\ub'.$$
Gronwall inequality gives 
$$\sum_{i=0}^{3}\int |\nabla^i\alphab|^2\leq \left(C\mathcal R_0+C(\mathcal O_0,\mathcal R)\epsilon^{\frac{1}{2}}\right)\exp\left(C(\mathcal O_0,\mathcal R_0, \mathcal G_0)\ub\right)\leq C(\mathcal O_0,\mathcal R_0, \mathcal G_0),$$
by choosing $\epsilon$ sufficiently small depending on $\mathcal R$.
Using this estimate for $\alphab$, we have
$$\mathcal R\leq C(\mathcal O_0,\mathcal R_0, \mathcal G_0)+C(\mathcal O_0,\mathcal R_0,\mathcal R, \mathcal G_0)\epsilon^{\frac{1}{2}},$$
from which the proposition follows.
\end{proof}
\section{Last Slice Argument and the End of Proof}\label{lastslice}
Consider the region $\mathcal D=\{0\leq u \leq \epsilon, 0\leq \ub\leq I\}$. This is the region in which we have proved a priori estimates. We would like to show that in fact the solution to the vacuum Einstein equations exists in this region. Let $t=u+\ub$. It is easy to see that $\nabla t$ is timelike by Propositions \ref{Omega} and \ref{b}. Define the level sets of $t$ to be $\Sigma_t$. By Rendall's Theorem, if the solution does not exist in the full region $\mathcal D$, we must have $t^*\in (0,I+\epsilon)$ such that $$t^*=\sup\{t:\mbox{the spacetime exists in }\mathcal D\cap \cup_{\tau\in[0,t)} \Sigma_\tau\}.$$
Let $g_t$, $k_t$ be the induced (Riemannian) metric and second fundamental form respectively on $\Sigma_t$. We will show that they converge in $C^\infty$ to $g_{t^*}$ and $k_{t^*}$ and that $g_{t^*}$ and $k_{t^*}$ satisfy the constraint equations. In order to show this, it suffices to show that all derivatives of $g_{t}$ is bounded uniformly in $L^2(\Sigma_t)$ for all $t<t^*$. Since we already have uniform estimates on $L^2(S)$ for the metric and the Ricci coefficients, it suffices to prove that $\nabla_3^i\nabla_4^j\nabla^k\Psi$ to be bounded in $L^2(\Sigma_t)$ uniformly in $t$ for $t<t^*$ for every $i,j,k$. The proof proceeds by induction on $i,j,k$.
Let $$E_{i,j,k}(t)=||\nabla_3^i\nabla_4^j\nabla^k\Psi||_{L^2(\Sigma_t)}.$$
The energy estimates in the previous section would then give
$$E_{i,j,k}(t)\leq C_{i,j,k},$$
where $C_{i,j,k}$ is a constant independent of $t$ for $t<t^*$ for $i=j=0, k\leq3$. 
Notice that using the estimates we have in the previous section, the Ricci coefficients can be controlled by 
\begin{equation}\label{horc}
||\nabla_3^{i}\nabla_4^{j}\nabla^{k}\psi||_{L^2(\Sigma_t)}\leq ||\nabla_3^{i}\nabla_4^{j}\nabla^{k}\psi||_{L^2(H_0)}+||\nabla_3^{i}\nabla_4^{j}\nabla^{k}\psi||_{L^2(\Hb_0)}+\sum_{i'\leq i, j'\leq j, k'\leq k}\sup_{t'\leq t}E_{i',j',k'}(t').
\end{equation}
This can be derived inductively by integrating the null structure equations as in the previous section. 
Now, using all the estimates in the previous section, together with (\ref{horc}), we can prove energy estimates using integration by parts as in the previous section. Notice that assuming the estimates in the previous section, the equations are always linear in the highest order term.
$$E_{i,j,k}(t)\leq (\sum_{i'<i,j'\leq j,k'\leq k} C_{i',j',k'})\int_0^t E_{i,j,k}(t') dt'.$$
$$E_{i,j,k}(t)\leq (\sum_{i'\leq i,j'< j,k'\leq k} C_{i',j',k'})\int_0^t E_{i,j,k}(t') dt'.$$
$$E_{i,j,k}(t)\leq (\sum_{i'\leq i,j'\leq j,k'<k} C_{i',j',k'})\int_0^t E_{i,j,k}(t') dt'.$$
By Gronwall's inequality, we would then have
$$E_{i,j,k}\leq C_{i,j,k},$$
where $C_{i,j,k}$ is a constant independent of $t$ for $t<t^*$ for all non-negative integers $i,j,k$.

On each spacelike hypersurface $\Sigma_t$, define also the coordinate $r=\ub-u$.
We will use the notation $(x^0,x^1,x^2,x^3)=(t,\th^1,\th^2,r)$.

Now on $\Sigma_{t^*}$, we define wave coordinates. For $a\in\{1,2,3\}$, let $g_{ab}=(g_{t^*})_{ab}$, $g_{00}=-1$, $g_{0a}=0$, $\partial_0 g_{ab}=2k_{ab}$. Then one can define $\partial_0 g_{a0}$ and $\partial_0 g_{00}$ in a way such that $\Gamma^\mu=g^{\nu\sigma}\gamma^\mu_{\nu\sigma}=0$. By virtue of the constraint equations on $\Sigma_{t^*}$, which is satisfied by continuity, we also have $\partial_0\Gamma^\mu=0$.

By standard theory of quasilinear wave equations, we have local existence for the reduced Einstein equation in the domain of dependence of $\Sigma_{t^*}$. By virtue of $\Gamma^\mu$ satisfying a wave equation and that $\Gamma^\mu=0$, $\partial_0 \Gamma^\mu=0$ on $\Sigma_{t^*}$, we know that in this coordinate system, $\Gamma^\mu=0$ everywhere, and in particular on the null hypersurface $\ub=t^*$. Define on this null hypersurface, the coordinates $(u,\th^1,\th^2)$, where $u=\frac{1}{2}(t^*-r)$. On the null hypersurface $H_0$, we prescribe the metric components as in Section \ref{local} except that we set $g_{34}=-2$ on $S_{0,t^*}$ (instead of $S_{0,0}$). The procedure in Section \ref{local} would allow us to find metric components and their derivatives on $H_0$ in a neighborhood of $S_{0,t^*}$ that is consistent with $\Gamma^\mu=0$ and $R_{\mu\nu}=0$. We now have two intersecting null hypersurfaces intersecting at the sphere $S_{0,t^*}$ and on both null hypersurfaces, the spacetime metric and its derivatives are given so as to be consistent with $\Gamma^{mu}=0$ and $R_{\mu\nu}=0$. We can thus solve the reduced Einstein equations as in section \ref{local} to obtain a vacuum Einstein spacetime.

Given $g_{\mu\nu}$ and $\partial_0 g_{\mu\nu}$, we can also solve the reduced Einstein equations backwards. Thus in the backwards domain of dependence, there exists an Einstein vacuum spacetime. Moreover, by the geometric uniqueness of the Cauchy problem in general relativity, this spacetime is diffeomorphic to the spacetime that we have obtained.
We now have a region in the spacetime containing a neighborhood of $\Sigma_{t^*}$ which is smooth in wave coordinates.

We have now extended the spacetime beyond the hypersurface $\Sigma_{t^*}$. It remains to show that we can change coordinates in a smooth fashion so that the spacetime can actually be extended beyond the hypersurface $\Sigma_{t^*}$ in canonical coordinates. To do so, we solve the Eikonal equations
\begin{equation}\label{finalu}
g^{\mu\nu}\partial_\mu u\partial_\nu u=0, g^{\mu\nu}\partial_\mu \ub\partial_\nu \ub=0,
\end{equation}
so that $u$ and $\ub$ agree with the the values of $u$ and $\ub$ in the canonical coordinates. Then solve for $\th^1$ and $\th^2$ by the transport equation
\begin{equation}\label{finalth}
\frac{\partial}{\partial \ub}\th^A=0.
\end{equation}
The equations (\ref{finalu}) and (\ref{finalth}) can be solved in a neighborhood of $\Sigma_{t^*}$ with smooth solutions $u,\ub,\th^1,\th^2$. By uniqueness of the equations, in a neighborhood to the past of $\Sigma_{t^*}$, these solutions agree with the canonical coordinate functions. We can then change to the $(u,\ub,\th^1,\th^2)$ coordinates. Therefore, we have extended the spacetime beyond the hypersurface $\Sigma_{t^*}$ in canonical coordinates. This would contradict the choice of $t^*$. Thus the solution to the vacuum Einstein equations exist in the full region in which we can prove a priori estimates. This proves Theorem \ref{mainthm}.
\section{Discussion}\label{discussion}
\subsection{Semilinear Equations in $3+1$ Minkowski Space}\label{semilinear}
In \cite{Rendall}, the local existence theorem is proved for general quasilinear wave equations. It is natural to ask whether our result extends to that general case. It is clear from our proof that the same idea would apply to semilinear wave equations with a null condition in $3+1$ dimensions as defined by Klainerman \cite{KlNull}. In the following, we work in $3+1$ Minkowski spacetime, where the metric is given by
$$m=-dt^2+\sum_{i=1}^3 dx_i^2.$$
We then define the null condition.
\begin{definition}
$\mathcal Q(\phi,\psi)$ is a null form if 
$$\mathcal Q(\phi,\psi)=A^{\mu\nu}\partial_\mu\phi\partial_\nu\psi,$$
and 
$$A^{\mu\nu}\xi_\mu\xi_\nu=0$$
whenever $\xi_\mu$ is null, i.e., $m(\xi,\xi)=0$.
\end{definition}
Define 
$$\Box_m=-\frac{\partial^2}{\partial t^2}+\sum_{i=1}^3\frac{\partial^2}{\partial x_i^2}.$$
Moreover, let
$$r=\sqrt{\sum_{i=1}^3 x_i^2}.$$
Then we have the following theorem:
\begin{theorem}
Consider a semilinear wave equation in Minkowski spacetime
$$\Box_m\phi=\mathcal Q(\phi,\phi).$$
Suppose $\phi$ is given and is smooth on the truncated outgoing null cone $t-r=-1$, $1\leq t+r\leq I$ and the truncated incoming null cone $t+r=1$, $-1\leq t-r <1-\delta$, with $\delta>0$. Then there exists $\epsilon=\epsilon(I,\delta, \mbox{initial data})$ such that a unique smooth solution exists in $\{-1 \leq t-r\leq -1+\epsilon,1 \leq t+r\leq I\}$ and $\{-1\leq t-r\leq 1-\delta,1\leq t+r\leq 1+\epsilon\}$
\end{theorem}
\begin{proof}
We prove this theorem in the spirit of our main theorem in this paper. We will prove the theorem in the region $\{-1 \leq t-r\leq -1+\epsilon,1 \leq t+r\leq I\}$. The region near the incoming cone can be treated analogously. In order to have consistent notations as above, we define
$$u=\frac 12 (t-r+1),\quad\ub=\frac 12 (t+r-1).$$
We also denote by $\nabla$ the angular derivatives. Note that $[\partial_u,\nabla]\sim[\partial_{\ub},\nabla]\sim[\nabla,\nabla]\sim \nabla$.
We name the norms in a way that one can compare with those in the main theorem.
$$\mathcal R=\sup_u(\sum_{i=0}^3||\nabla^i\partial_{\ub}\phi||_{L^2(H_u)}+\sum_{i=1}^4||\nabla^i\phi||_{L^2(H_u)})+\sup_{\ub}(\sum_{i=0}^3||\nabla^i\partial_u\phi||_{L^2(\Hb_{\ub})}+\sum_{i=1}^4||\nabla^i\phi||_{L^2(\Hb_{\ub})}).$$
Define also the initial data norms by
$$\mathcal R_0=\sum_{i=0}^3||\nabla^i\partial_{\ub}\phi||_{L^2(H_0)}+\sum_{i=1}^4||\nabla^i\phi||_{L^2(H_0)}+\sum_{i=0}^3||\nabla^i\partial_u\phi||_{L^2(\Hb_{0})}+\sum_{i=1}^4||\nabla^i\phi||_{L^2(\Hb_{0})},$$
$$\mathcal O_0=\sup_{\ub}(\sum_{i=0}^2||\nabla^i\partial_{\ub}\phi||_{L^\infty(S_{0,\ub})}+\sum_{i=0}^3||\nabla^i\phi||_{L^\infty(S_{0,\ub})})+\sup_{u}(\sum_{i=0}^2||\nabla^i\partial_u\phi||_{L^\infty(S_{u,0})}+\sum_{i=0}^3||\nabla^i\phi||_{L^\infty(S_{u,0})}),$$

The most important property about the null condition is that when we write the derivatives of $\phi$ in terms of $\partial_{\ub}\phi$, $\partial_u\phi$ and $\nabla\phi$, the nonlinearity does not contain the term $(\partial_{\ub}\phi)^2$ or $(\partial_u\phi)^2$.

We first need some estimates on $\phi$. It is clear that if $||\partial_u\phi||_{L^\infty(S_{u,\ub})}<\infty$, then $\epsilon$ can be chosen sufficiently small so that $$||\phi||_{L^\infty(S_{u,\ub})}\leq 2\mathcal O_0.$$

Next, we assume that $\mathcal R<\infty$ and we would like to use transport type equations to prove estimates of $\phi$ up to three derivatives in $L^2(S)$.
For $\nabla^k\phi$, we will simply estimate by the $L^2(\Hb)$ norms of $\partial_u\nabla^k\phi$. For $\partial_u\nabla^k\phi$ and $\partial_{\ub}\nabla^k\phi$, we will use the equation, which can be
rewritten as
$$\partial_u\partial_{\ub}\phi+\nabla^2\phi+\partial\phi=(\partial_u\phi+\nabla\phi)(\partial_{\ub}\phi+\nabla\phi).$$
For higher derivatives, we take angular derivatives of this equation and make the necessary commutations.
It is then clear that by choosing $\epsilon$ to be sufficiently small,
\begin{equation}\label{goodterms}
\sum_{i=0}^2||\partial_{\ub}\nabla^i\phi||_{L^2(S_{u,\ub})}+\sum_{i=0}^3||\nabla^i\phi||_{L^2(S_{u,\ub})}\leq C(\mathcal O_0).
\end{equation}
In estimating the terms with $\partial_u$ derivatives, the key is to notice that due to the null structure, we can use Gronwall inequality to derive
\begin{equation}\label{badterms}
\sum_{i=0}^2||\partial_{u}\nabla^i\phi||_{L^2(S_{u,\ub})}\leq C(\mathcal O_0, \mathcal R,I).
\end{equation}
Notice that we also have the corresponding $L^4(S)$ bounds up to $2$ derivatives and $L^\infty(S)$ bounds up to $1$ derivative by Sobolev Embedding.

Finally, we have the following energy estimates:
$$\mathcal R^2\leq \mathcal R_0^2+\sum_{\psi_1\in\{\partial_{\ub}\phi,\nabla\phi\},\psi\in\{\partial_{\ub}\phi,\partial_u\phi,\nabla\phi\}}\int_{D_{u,\ub}}\sum_{i=0}^3\nabla^i\psi_1\sum_{j=0}^3\nabla^j(\psi\psi)+\sum_{i=0}^3\nabla^i\psi\sum_{j=0}^3\nabla^j(\psi_1\psi)+\sum_{i,j=0}^3\nabla^i\psi\nabla^j\psi.$$
This can be derived by integration by parts. For the nonlinear terms, the two factors with the highest derivatives are controlled in $L^2(H_u)$ or $L^2(\Hb_{\ub})$. The remaining factor is necessarily lower order and thus are controlled using (\ref{goodterms}) and (\ref{badterms}). When at least one of the highest order term can be controlled in $L^2(H_u)$, we can make use of the integration in the $u$ direction to gain a smallness factor $\epsilon^\alpha$. The danger arises when both of these terms have $\partial_u$ derivative and can only be controlled in $L^2(\Hb_{\ub})$. Nevertheless, as seen in the above schematic equation, the structure of the equation would dictate that the lower order factor to be controllable by (\ref{goodterms}). Hence, such terms are linear in $\mathcal R^2$ and can be handled by Gronwall's inequality. The last term in the schematic equation is also linear in $\mathcal R^2$ and can be treated by Gronwall's inequality. Thus we have
$$\mathcal R^2\leq \mathcal R_0^2+\epsilon^{\frac{1}{2}}C(\mathcal O_0,\mathcal R).$$
We can thus close the bootstrap assumption by choosing $\epsilon$ sufficiently small.
\end{proof}
\subsection{Counterexample without a Null Condition}\label{counterexample}
However, without a null condition, the theorem no longer holds. This is in contrast with Rendall's theorem, which holds for general quasilinear wave equations. To see that the theorem fails without a null condition, we can consider the scalar equation
$$\Box_m\phi=(\partial_u\phi)^2.$$
Prescribe $\phi=0$ on $H_0$ and $\phi$ spherically symmetric on $\Hb_0$ such that $\partial_u\phi<-C$ for some large positive $C$ to be determined in a neighborhood of $S_{0,0}$. With $u=\frac 12 (t-r+1),\quad\ub=\frac 12 (t+r-1)$, we would like to show that for every $\epsilon$, the solution $\phi$ fails to exist in $0\leq \ub\leq 1, 0\leq u\leq \epsilon$. To see this, fix $\epsilon$. Assume on the contrary that we have a $C^1$ solution in the region. Then, for any small $\delta$, we have $|\partial_{\ub}\phi|\leq \delta$ by choosing $\epsilon$ smaller if necessary. Now, we can write the equation as follows:
$$\partial_{\ub}\partial_u\phi=-\frac 2 r \partial_r\phi-(\partial_u\phi)^2\leq -(\partial_u\phi)^2-\frac 1 r \partial_u\phi+\delta=-(\partial_u\phi+\frac{1}{2r^{\frac{1}{2}}})^2+\frac 1 r+\delta.$$
This implies, in the region $0\leq \ub\leq 1, 0\leq u\leq \epsilon$,
$$\partial_{\ub}(\partial_u\phi+\frac{1}{2r^{\frac{1}{2}}})\leq -(\partial_u\phi+\frac{1}{2r^{\frac{1}{2}}})^2+\frac 1 r+\frac{2}{r^{\frac 32}}+\delta\leq -(\partial_u\phi+\frac{1}{2r^{\frac{1}{2}}})^2+4.$$
This is an ODE in $\partial_u\phi+\frac{1}{2r^{\frac{1}{2}}}$ for each fixed $\ub$, which clearly blows up before $\ub = 1$ if $C$ is chosen to be sufficiently large. Thus a $C^1$ solution does not exist in $0\leq \ub\leq 1, 0\leq u\leq \epsilon$ for any $\epsilon$.

\subsection{Regularity}\label{regularity}
It is a natural question to ask what regularity is required for our main Theorem. In other words, it would be interesting to understand whether $\epsilon$ can be chosen to depend on a weaker norm of the initial data. In the proof, we have used 3 angular derivatives of the curvature in $L^2(H)$ and $L^2(\Hb)$. Since in this paper we are interested in the existence in the category of smooth solutions, we did not optimize the regularity used in the argument. Nevertheless, we would like to point out that it is possible to reduce to only 1 derivative of the curvature (now including also the $\nabla_3$ and $\nabla_4$ derivatives) using only elliptic estimates and trace estimates. We now sketch a proof.

Define accordingly 
$$\mathcal R= \sum_{i=0}^{1}\sup_u\sup_{\Psi\in\{\alpha,\beta,\rho,\sigma,\betab\}}||\nabla^i\Psi||_{L^2(H_u)}+\sup_{\ub}\sup_{\Psi\in\{\beta,\rho,\sigma,\betab,\alphab\}}||\nabla^i\Psi||_{L^2(\Hb_{\ub})}+\sup_u||\nabla_4\alpha||_{L^2(H_u)}+\sup_{\ub}||\nabla_3\alphab||_{L^2(\Hb_{\ub})},$$
$$\mathcal R(S)=\sup_{u,\ub}||\alpha,\beta,\rho,\sigma,\betab||_{L^4(S_{u,\ub})},$$
Assume as bootstrap assumption, instead of only $\mathcal R<\infty$ and $\mathcal R(S)<\infty$, also that $||\nabla^2\psi||_{L^2(H_u)}<\infty,$ $||\nabla^2\psi||_{L^2(\Hb_{\ub})}<\infty$. First, we try to estimate the $||\psi||_{L^\infty(S)}$ and $||\nabla\psi||_{L^2(S)}$. $||\psi||_{L^\infty(S)}$ would be estimated as in Proposition \ref{RicciLinfty} except that we now need to estimate
$$\sup_{\th^1,\th^2} \int_0^u \Psi^2 du',\quad\sup_{\th^1,\th^2} \int_0^u (\nabla\eta)^2 du',\quad \sup_{\th^1,\th^2} \int_0^{\ub} \Psi^2 d\ub'.$$
These in turn can be estimated by $\mathcal R$ and $||\nabla^2\eta||_{L^2(\Hb)}$ by the trace estimates. (Here is where the assumption for $\nabla_3\alphab$ and $\nabla_4\alpha$ are needed.) See \cite{KlRo}. The estimates for $||\nabla\psi||_{L^2(S)}$ can be estimated as in Proposition \ref{RicciL2}. Notice that the estimates for $||\trch,\trchb,\chih,\chibh,\etab,\omega||_{L^\infty(S)}$ and $||\nabla(\trch,\trchb,\chih,\chibh,\etab,\omega)||_{L^2(S)}$ are still dependent only on the initial data. $||\eta,\omega||_{L^\infty(S)}$ and $||\nabla(\eta,\omegab)||_{L^2(S)}$, on the other hand, depend on the initial data and $\mathcal R$.

Notice also that the $L^\infty$ control we obtained for the Ricci coefficients is sufficient to run the argument in Section \ref{geometry} to control the geometry and to show the necessary Sobolev Embedding and the estimates for transport equations.

The next step would be to show the boundedness of $\mathcal R(S)$ independent of $\mathcal R$. This is done via the codimension 1 trace estimates as in Proposition \ref{RS}, i.e., we can first prove the $L^2(S)$ estimate and use that to gain a smallness constant in the codimension 1 trace estimates to conclude that $\mathcal R(S)\leq C(\mathcal O_0,\mathcal R_0,\mathcal G_0)$. Once this is achieved, we can also control $||K||_{L^2(S)}$ independent of $\mathcal R$ so that we can apply elliptic estimates. In this setting, we can also control $\alphab$ in $L^4(S)$ because we have the estimates for $\nabla_3\alphab$ in the norm $\mathcal R$.

We now need to estimate $||\nabla^2\psi||_{L^2(H)}<\infty,$ $||\nabla^2\psi||_{L^2(\Hb)}<\infty$ assuming $\mathcal R<\infty$. This is done via systems of transport and elliptic equation. One can find combinations of $\Theta=\nabla\psi+\Psi$ such that $\nabla_3\Theta=\psi\nabla\psi+\psi^3+\psi\Psi$ or $\nabla_4\Theta=\psi\nabla\psi+\psi^3+\psi\Psi$. One can then solve for the necessary components of $\nabla\psi$ from these $\Theta$. See \cite{Chr},\cite{CK},\cite{KN},\cite{KlRo}, etc. Once this is done, we can use Proposition \ref{trace} to get estimates for $||\nabla\psi||_{L^4(S)}$. 

Now, we prove energy estimates for the curvature. This proceeds in similar way as in Proposition \ref{finalcurvbound} except that now we also need estimates for $\nabla_3\alphab$ and $\nabla_4\alpha$, which we ignore for the moment. The most important observation is that the $L^\infty(S)$ bounds for $\trch,\trchb,\chih,\chibh,\etab,\omega$ depends only on the initial data. We would also have terms that look like
$$\int_{D_{u,\ub}}\nabla\alphab \nabla\psi \alphab.$$
To control this term, notice that since $\alphab$ can be controlled in $L^4(S)$ independent of $\mathcal R$, we can estimate this terms by $\epsilon \mathcal R$.

In order to close these estimates for $\nabla_3\alphab$ and $\nabla_4\alpha$, we also need estimates for $||\nabla_3\psi||_{L^4(S)}$ and $||\nabla_4\psi||_{L^4(S)}$. For $\psi$ satisfying $\nabla_3\psi$ equation, the $||\nabla_3\psi||_{L^4(S)}$ estimate can be obtained by directly estimating $\nabla_3\psi=\Psi+\nabla\psi+\psi\psi$ since all the terms on the right hand side is in $L^4(S)$. Similarly, for $\psi$ satisfying a $\nabla_4\psi$ equation, we can obtain the $||\nabla_4\psi||_{L^4(S)}$ estimate easily. This leaves $\nabla_3\eta$, $\nabla_4\etab$, $\nabla_3\omegab$ and $\nabla_4\omega$. To estimate their $L^4(S)$ norms, we resort to the codimensional 1 trace estimates. For definiteness, we indicate how this can be achieved for $\nabla_3\eta$. All other quantities can be estimated analogously. In order estimate $\nabla_3\eta$, we need to estimate $\nabla_4\nabla_3\eta$, $\nabla\nabla_3\eta$ and $\nabla_3\eta$ is $L^2(H)$. $\nabla_3\eta$ has been estimated. $\nabla_4\nabla_3\eta$ is estimated directly after differentiating the equation for $\nabla_4\eta$ by $\nabla_3$. Finally, $\nabla\nabla_3\eta$ using the $\Theta$ quantity associated to $\nabla\eta$, which looks like $\Theta=\nabla\eta+\Psi$ and satisfies a equation of the form $\nabla_4\Theta=\psi\nabla\psi+\psi^3+\psi\Psi$. Differentiating this equation by $\nabla_3$ and commuting $\nabla_3$ and $\nabla_4$, we get a transport equation for $\nabla_3\Theta$, using which we can estimate $\nabla_3\Theta$ in $L^2(S)$. Then we can use elliptic estimates to put $\nabla\nabla_3\eta$ in $L^2(H)$. A similar strategy would then allow use to estimate all first derivatives of $\psi$ in $L^4$. Notice that these estimates depend on $\mathcal O_0$, $\mathcal R_0$ and $\mathcal R$.

We finally move to energy estimates for $\nabla_3\alphab$ and $\nabla_4\alpha$. We need to commute the Bianchi equations for $(\nabla_4\alphab,\nabla_3\betab)$ with $\nabla_3$ and similarly commute the Bianchi equations for $(\nabla_4\beta,\nabla_3\alpha)$ with $\nabla_4$. The most dangerous term is now of the form
$$\int_{D_{u,\ub}} \nabla_3\alphab \psi\nabla_3\alphab.$$
As in Proposition \ref{finalcurvbound}, we note that the components $\psi$ that are allowed in this expression must satisfy a $\nabla_3$ equation and can thus be controlled in $L^\infty$ independent of $\mathcal R$. We can then conclude with Gronwall inequality.

\bibliographystyle{hplain}
\bibliography{NewLocal}

\end{document}